\documentclass[%
 reprint,
 amsmath,amssymb,
 aps,
]{revtex4-2}

\usepackage{graphicx}
\usepackage{dcolumn}
\usepackage{bm}
\usepackage{soul,xcolor}
\usepackage{physics}
\usepackage{amsmath}
\usepackage{siunitx}
\usepackage{tabularx}
\usepackage{array}
\usepackage{float}
\usepackage{booktabs}
\usepackage{array,multirow}
\usepackage{hyperref}
\usepackage{orcidlink}

\hypersetup{
    colorlinks=true,
    linkcolor=blue,
    citecolor=blue,
    filecolor=blue,
    urlcolor=blue
}

\newcolumntype{Y}{>{\raggedright\arraybackslash}X}
\renewcommand{\a}{{\hat{a}}}

\newcommand{\ad}{{\hat{a}^{\dagger}}}

\newcommand{\Erec}{E_{R}} 
\newcommand{\HH}{\hat{\mathcal{H}}}
\newcommand{\FF}{\mathcal{F}}

\newcommand{\LL}{\hat{\hat{\mathcal{L}}}}

\newcommand{\lcav}{\ell_{\rm cav}} 
\def\lconv{{\ell_{\rm conv}}}
\def\lc{{\ell_{\rm c}}} 
\newcommand{\mSr}{{m_{\rm Sr}}}

\newcommand\oo{\hat{O}}

\def\transclock{{$^1$S$_0$}\,$\rightarrow$\,{$^3$P$_0$}~}
\def\transred{{$^1$S$_0$}\,$\rightarrow$\,{$^3$P$_1$}~}
\newcommand{\Uconv}{U_{\rm conv}}   
\newcommand{\Veff}{{V_{\rm eff}}}
\newcommand{\vconv}{{v_{\rm conv}}} 

\newcommand{\gamdep}{\gamma_{\rm dep}}

\newcommand{\sig}{{\hat{\sigma}}}

\begin{document}

\preprint{APS/123-QED}

\title{Modeling of a continuous superradiant laser on the sub-mHz \transclock transition in neutral strontium-88}
\author{Swadheen Dubey\,\orcidlink{0000-0001-5622-4706}} \author{Georgy A. Kazakov\,\orcidlink{0000-0003-4110-2380}}\email{kazakov.george@gmail.com}
    \affiliation{Atominstitut, TU Wien, Stadionallee 2, 1020 Vienna, Austria}
\author{Benedikt Heizenreder\,\orcidlink{0000-0001-5611-4144}}
\author{Sheng Zhou\,\orcidlink{0000-0002-8230-6577}}
\author{Shayne Bennetts\,\orcidlink{0000-0001-8684-6194}}
\author{Stefan~Alaric~Sch\"{a}ffer\,\orcidlink{0000-0002-5296-0332}}\thanks{Present address: NQCP, Niels Bohr Institute, University of Copenhagen}
\author{Ananya Sitaram\,\orcidlink{0000-0002-6548-4515}}
\author{Florian Schreck\,\orcidlink{0000-0001-8225-8803}}
\email{Modeling88SrSuperradiance@strontiumBEC.com}
     \affiliation{Van der Waals - Zeeman Institute, Institute of Physics, University of Amsterdam, Amsterdam, the Netherlands}

\date{\today}
\collaboration{MoSaiQC Collaboration}

\date{\today}

\begin{abstract}
Continuous superradiance using a narrow optical transition has the potential to improve the short-term stability of state-of-the-art optical clocks. Even though pulsed superradiant emission on a ${\rm mHz}$ linewidth clock transition has been shown, true continuous operation, without Fourier limitation, has turned out to be extremely challenging. The trade-off between maintaining a high atomic flux while minimizing decoherence effects presents a significant obstacle. Here, we discuss the design of a machine that could overcome this problem by combining a high-flux continuous beam of ultracold strontium atoms with a bowtie cavity for the generation of superradiant lasing. To evaluate the feasibility of our design, we present simulation results for continuous high-efficiency cooling, loading, and pumping to the upper lasing state inside the bowtie cavity. We then present two different models for simulating the generated superradiant field by taking into account position-dependent shifts, collisional decoherence, light shifts, and atom loss. Finally, we estimate a laser linewidth of less than $100~{\rm mHz}$, limited by atom number fluctuations, and resulting in an output power of hundreds of ${\rm fW}$. 
\end{abstract}

\maketitle            
\section{Introduction}
\label{sec:intro}
Optical atomic clocks are among the most precise devices in the world, reaching instabilities of $6.6 \times 10^{-19}$ after 1 hour of averaging~\cite{Oelker19}. 
State-of-the-art optical clocks operate \textit{passively}, meaning the frequency of the local oscillator is intermittently compared with the frequency of a narrow clock transition in an ensemble of trapped atoms or ions. The local oscillator is typically a laser pre-stabilized to a high finesse ultrastable macroscopic cavity. One of the main factors limiting the short-term stability of optical clocks is the instability of this macroscopic cavity, which is susceptible to thermal fluctuations~\cite{kedar_frequency_2023,yu_excess_2023}.

Conventional lasers operate in the {\em good-cavity regime}, where the cavity loss rate $\kappa$ is much less than the frequency gain bandwidth $\Gamma_\mathrm{gain}$ of the gain medium. The phase information of such a laser is encoded in the light field. Here, the output frequency is determined by the frequency $\nu_\mathrm{cavity}$ of a resonant cavity mode, which in turn follows the fluctuations of the cavity length. In contrast, in {\em bad-cavity lasers}, $\Gamma_{\rm gain} \ll \kappa$, and the output frequency is determined primarily by the transition frequency of the gain medium, making it almost completely insensitive to fluctuations of the cavity length. The remaining shift can be described by the ``cavity pulling'', which is reduced by a factor $\Gamma_\mathrm{gain}/\kappa$ compared to a {\em good-cavity laser} \cite{Chen09}. This opens the possibility of creating an {\em active frequency standard} by combining a ``bad" cavity with an extremely narrow bandwidth gain medium, such as forbidden transitions in alkaline-earth atoms~\cite{Yu08,Meiser09}.
Because the linewidth of such a laser is mainly determined by the gain medium itself, spontaneous decay of the excited state, pumping field, etc., will lead to phase diffusion of the collective atomic dipole, which limits the achievable performance. Nevertheless, a sub-millihertz level laser in the optical frequency domain is within reach \cite{Kolobov93, Meiser09, Chen09}.

A special case of a bad-cavity laser is a so-called {\em superradiant laser}~\cite{Jaeger21}. In this type of laser, the cavity linewidth $\kappa$ is large in comparison to the cavity-mediated collective decay rate of the gain medium.
Superradiance refers to the collective emission of light, enhanced by mutual synchronization of radiating dipoles that interact with each other \cite{Gross82}. 
Superradiance has been initially predicted for the case of multiple emitters localized in a region smaller than the radiation wavelength \cite{Dicke54}. Here, they are prepared in the excited state and then emit a single bell-shaped light pulse whose peak intensity scales as $N^2$, where $N$ is the number of emitters. Counterintuitive to this picture, superradiance can also be observed if the emitters are in a dilute gas, but are synchronized by a shared cavity mode. More explicitly, the strong collective coupling created by an overdamped cavity mode is used to synchronize them \cite{Norcia16A, Laske19}. 
In addition, if the inversion of the gain can be maintained by some pumping mechanism, we can leave the pulsed regime and achieve a continuously operating superradiant laser.

True superradiant lasing is difficult to realize experimentally, so many experiments work in the {\em superradiant crossover regime}~\cite{Norcia16,Schaeffer20, Gogyan20}. In this regime, the cavity linewidth $\kappa$ is larger than $\Gamma_\mathrm{gain}$, but smaller than the collectively enhanced decay rate of the gain medium. In pulsed operation, the peak intensity of the light pulse scales linearly with the number of atoms \cite{Schaeffer20, Gogyan20}, in contrast to the quadratic scaling expected in the ``true" superradiant regime. In addition, reabsorption of the intracavity photons and subsequent reemission can lead to oscillations in the intensity of the light pulse, as was observed in Ref.~\cite{Norcia16}. These oscillations can lead to instability in a continuously  operated system, making the crossover regime not ideal for an active frequency standard \cite{Haken1981Light, kazakov171}.

To date, a continuously operating active optical clock has still not been created, although important steps towards this goal have been taken. In particular, pulsed superradiant lasing using the extremely narrow optical \transclock transition has been observed with $\rm ^{87}Sr$ atoms~\cite{Norcia16A, Norcia18, Norcia2018B}. In addition, quasi-continuous laser operation has been demonstrated in the superradiant crossover regime on the ${\rm ^1S_0\rightarrow ^3P_1}$ transition of ${\rm ^{88}Sr}$ \cite{Norcia16, Schaeffer20}.
In such a system, a sub-natural linewidth has been achieved by optical repumping inside the emission zone during the emission \cite{kristensen_subnatural_2023}. Unavoidably, this leads to light shifts and extra dephasing, which can be reduced, but not totally eliminated \cite{Hotter21repump}.
Moreover, loss of atoms from the cavity makes truly continuous operation in this kind of system impossible.

To overcome these limitations and achieve continuous operation, we can continuously supply new atoms in the upper lasing state into the emission zone instead of repumping the same sample of atoms. For this, a moving magic-wavelength optical lattice, or an optical conveyor, can be used, which reduces perturbations on the clock transition.
Such an optical conveyor can be formed by two counter-propagating running waves with slightly different frequencies excited within a ring cavity. 
The optical conveyor potential confines the atoms deep in the Lamb-Dicke regime along the cavity axis, which provides coherent interaction on the clock transition with the resonant running-wave mode of the same cavity.
Such a configuration, in comparison to one in which the conveyor runs perpendicular to the cavity axis, allows for increased atom-cavity interaction time, while keeping the density of the atoms moderate~\cite{Kazakov21EFTF}. A similar configuration has also been proposed to realize continuously operating passive optical lattice clocks with Ramsey interrogation ~\cite{katori_longitudinal_2021}. 
One of the main experimental challenges in such a system is the continuous loading of high flux of cold atoms into a ring cavity. Recent experiments have demonstrated steps towards this goal, namely loading and moving of $\mu$K cold atoms through a ring cavity~\cite{Cline2022, Schaefer2024, Takeuchi_2023, Katori24}. These promising results open the door to reaching truly continuous superradiance on an extremely narrow transition in neutral atoms soon. At the same time, the challenges associated with a truly continuous system demand for in-depth simulations of the main decoherence effects and systematics in such systems in order to assess the expected performance of a continuous superradiant laser.

In this article, we describe our apparatus at the University of Amsterdam, which combines a truly continuous ultracold strontium atom source with a ring cavity in order to produce continuous superradiance on the \transclock transition in the future.
We simulate the continuous loading of ${\rm ^{88}Sr}$ atoms into the ring cavity, the pumping process to the upper lasing state to create a continuously inverted gain medium, and finally, the generation of superradiant emission into a ring cavity mode~\cite{Kazakov21EFTF, Cline22}. We also theoretically study the properties of the generated light field and examine the main systematics in the system.

\section{Design of Apparatus}
\label{sec:exp}
\begin{figure*}
    \centering
    \includegraphics[width=1\textwidth]{"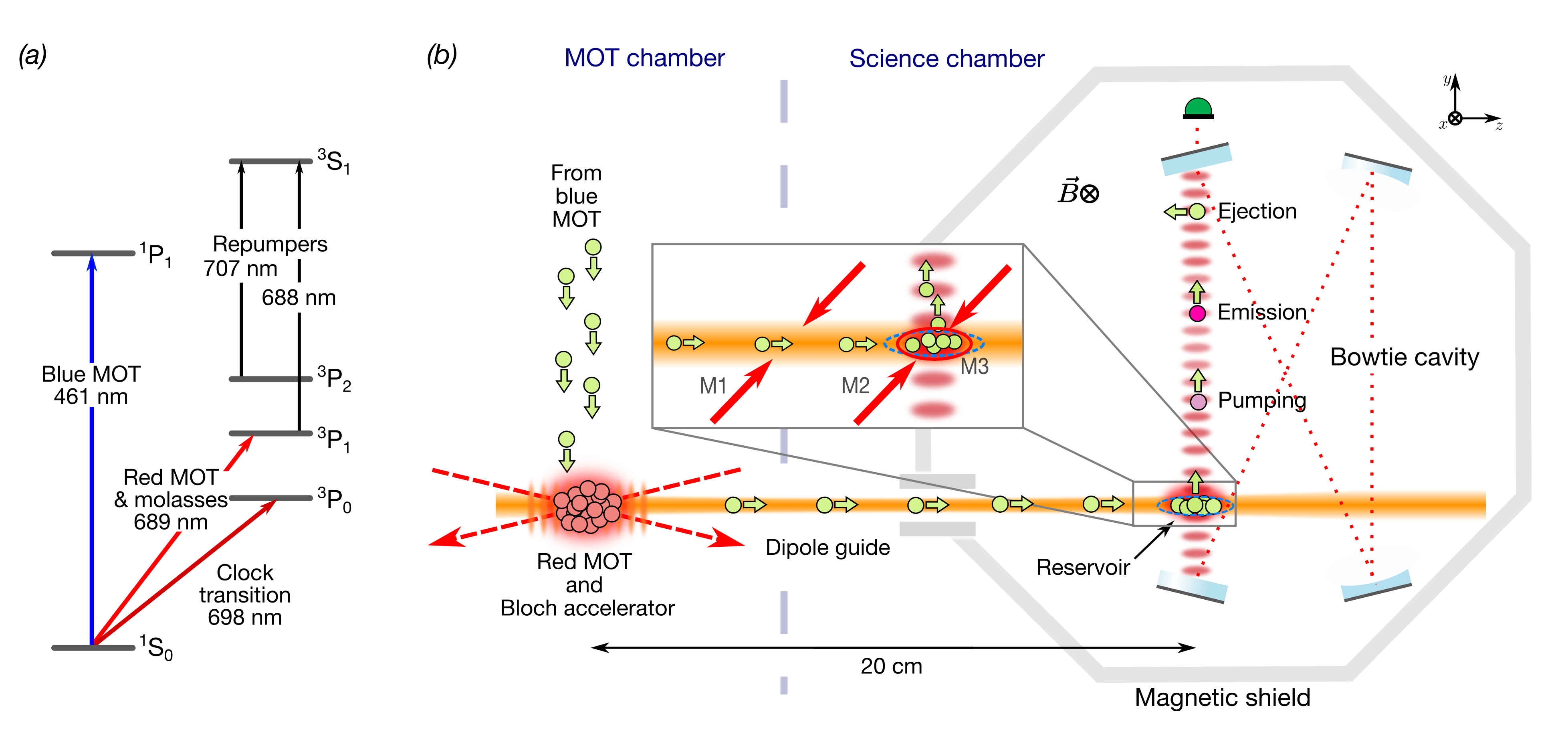"}
    \caption{
    The architecture of the 
    apparatus. Atoms falling down from a 2D blue MOT (not shown) are continuously collected and cooled in a red MOT. From there, they are loaded into a dipole guide formed by an incoherent $1070\,$nm laser beam. The guide beam is overlapped with a shallow-angle lattice (dashed arrows), which can be used as a Bloch accelerator. The atoms are transported $\sim$\SI{20}{\,\cm} from the red MOT chamber to the magnetically shielded science chamber. In the science chamber, the atoms are initially decelerated using a retro-reflected molasses beam operating on the \SI{7.5}{\kilo\Hz} \transred transition (M1). They are then further cooled into a reservoir dipole trap (denoted by blue dashed oval, formed by crossing of the guide beam with an extra $1070\,$nm  beam propagating into the $(x,z)$-plane and tilted by $5^\circ$ with respect to the $z$-axis) by the second (M2) and third (M3) set of red retro-reflected molasses beams where (M3) is orthogonal to the $(y,z)$-plane.
    Next, they are loaded into the magic wavelength optical conveyor lattice. The optical conveyor moves upwards through a pumping zone, where atoms are pumped to the excited $^3$P$_0$ state. Finally, they are transported through a well-controlled emission zone, where superradiant emission into the cavity mode can occur. Here, the bias magnetic field is generated by a set of Helmholtz coils inside the shield (not shown). After the emission zone, atoms are ejected by a push beam to avoid coating the surface of the cavity mirrors.}
\label{fig:SC}
\end{figure*}
Our apparatus uses a steady-state narrow-line magneto-optical trap (MOT) operating on the {$^1$S$_0$}\,$\rightarrow$\,{$^3$P$_1$}~ transition, which provides a continuous $\mu$K-temperature source of strontium. We employ a very similar design to the one described in Refs.~\cite{HighPSDMOT2017,Gonzalez2021,Chen2022}, in which we separate the laser cooling and trapping stages in space rather than in time. Here we can achieve continuous narrow-line MOT loading with a rate of up to about $10^7~{\rm s^{-1}}$ for $\rm ^{87}Sr$ and $10^8~{\rm s^{-1}}$ for $\rm ^{88}Sr$. 

Once we have a continuous $\mu$K source of atoms, the atoms must be coupled to each other to enable superradiant emission. 
To create the coupling field, we use a high-finesse ring cavity in a bowtie configuration.
Emission on the \transclock transition requires isolation from stray photons and the inhomogeneous magnetic fields of the continuously operating MOTs, which can cause decoherence and introduce inhomogeneous broadening.
We satisfy these requirements by separating the science chamber that houses the ring cavity from the red MOT chamber with a differential pumping tube and enclose the science chamber in a magnetic shield to prevent interference from the Zeeman slower and the MOTs' magnetic fields. To ensure a uniform magnetic field along the cavity mode, especially in the region where we expect superradiant emission, we used COMSOL to evaluate the magnetic field uniformity, including the Helmholtz magnetic field coils around the science chamber, the Zeeman slower and MOT magnetic fields, as well as the magnetic shields for the science chamber. The magnetic field inside the shielding is mainly determined by the two Helmholtz coils located inside the shielding.

The distance between the red MOT and the optical lattice is $20\,$cm, and the atoms must be transported and loaded into the lattice in a continuous fashion, without significant heating. Thus, we choose to transport the atoms using a combination of a Bloch accelerator~\cite{peik_bloch_1997} and a $200\,$W, 150$\,\mu$m-waist dipole guide beam with 1070-nm wavelength. The focus of the dipole guide beam lies at the halfway point between the red MOT and the optical lattice. The Bloch accelerator, created by the interference of two independent off-resonant laser beams aligned at a shallow angle with respect to each other, overlaps the red MOT. By detuning the frequency of one beam from the other, we can create a moving potential to accelerate the atoms out of the MOT and along the dipole guide beam, as shown in Fig.~\ref{fig:SC}. Atoms then travel along the dipole guide with a velocity of up to $50~{\rm cm/s}$, while being supported against gravity.

As the atoms arrive in the science chamber, they are decelerated by a 689-nm molasses beam and captured in a crossed dipole trap, which we call the ``reservoir". The reservoir trap consists of another 1070-nm beam propagating in the $(x,z)$-plane, tilted by $5^\circ$ with respect to the $z$-axis, and is also overlapped with the moving optical lattice. Two additional red molasses beams, depicted by a red line and red oval in Fig.~\ref{fig:SC}, aid in cooling the atoms into the reservoir. The main purpose of reservoir trap is to create a well-defined potential in which to store cold atoms and to increase the loading efficiency into the moving optical lattice.
The vertically oriented lattice, power-enhanced by the bowtie cavity, then acts as a conveyor, moving the atoms from the reservoir upwards through the pumping region, where they are optically pumped to the $^3$P$_0$ state~(Sect.~\ref{sec:pump}). Once in the excited clock state, the atoms continue traveling vertically to a well-controlled emission zone, where superradiant emission into the resonant cavity mode can occur~(Sect.~\ref{sec:las}).
Near the top of the cavity, atoms are ejected from the conveyor using a ``push" beam resonant with a cycling transition so they do not coat the cavity mirrors. 

The crux of the design is the bowtie cavity, which creates not only the conveyor lattice, which can only be achieved in a ring cavity design, but also the coupling field necessary to enable superradiant emission. We opt for a bowtie cavity design instead of a three-mirror ring cavity configuration since the close to normal angles between the cavity mode and the mirrors in a bowtie cavity lead to minimal optical aberrations of the cavity mode. A bowtie design also provides a compact cavity format, and requires only optical access from one side.
The overall dimensions of the cavity are $50\,$mm in length and $13\,$mm in width, and it is designed to have a free spectral range of $1.5\,$GHz and a finesse of about 50,000 at 698~nm and 2000 at $813\,$nm. We use mirrors with an ultra high reflectivity coating for $698\,$nm, keeping one outcoupler mirror with a slightly higher transmission, which will allow us to increase the outcoupled superradiant laser output, while maintaining the desired finesse. In the interrogation zone, the minimum waist of the moving lattice is around 140\,${\rm \mu m}$ and the $689\,$nm cavity mode will be $\approx10\,{\rm \mu m}$ smaller. To create a resonant cavity mode with the \transclock transition, the length of the cavity can be tuned using a piezo stack attached to one of the flat mirrors.

\begin{figure*}
    \centering
    \includegraphics[width=0.99\textwidth]{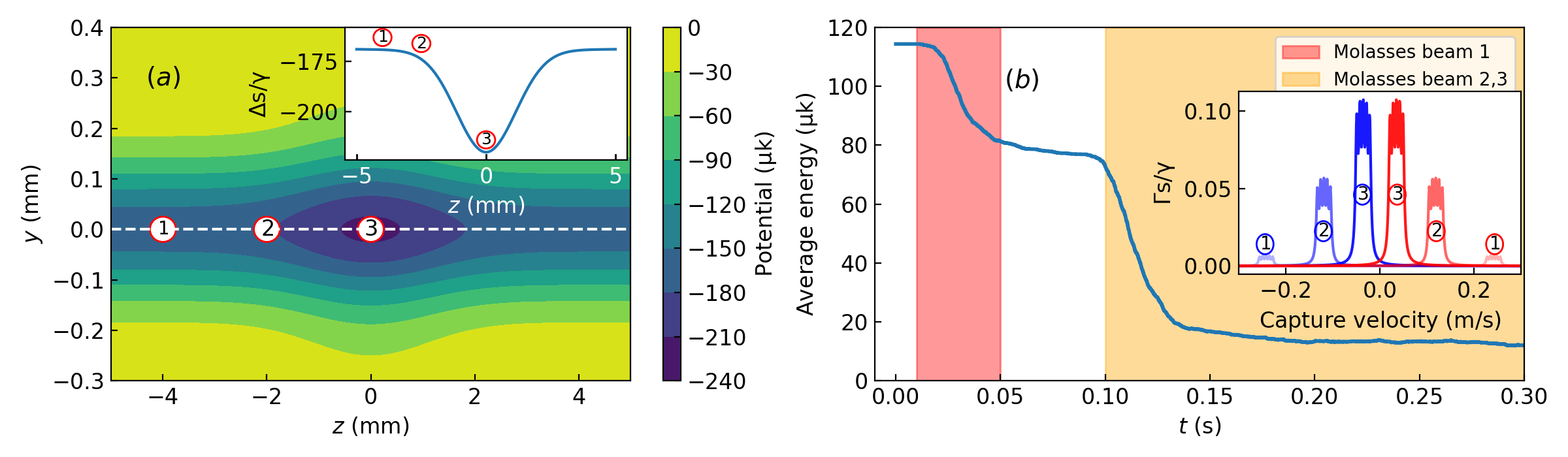}
    \caption{(a) Potential of reservoir and dipole guide in $(y,z)$-plane. The inset shows the differential light shift in the reservoir for the ${\rm ^1S_0} - {\rm ^3P_1}$ cooling transition, along the $z$-axis. (b) Cooling dynamics of atoms with an incoming velocity of 10\,cm/s, where the interaction time for each molasses beam is marked by the red or orange shading. The first molasses beam (M1, red shading), slows atoms to 1\,cm/s. Shortly afterwards, the atoms enter the reservoir region, where other molassess beams (M2 \& M3, orange shading) slow and cool them inside the reservoir. Finally they can reach $\mu$K-temperatures. The inset shows how the capture velocity changes with atomic position inside the reservoir. To avoid heating the frequencies of M2 and M3 are chosen such that there is no unwanted interaction with cold atoms in the center of the reservoir.}
    \label{fig:trap}
\end{figure*}

Our apparatus can accommodate experiments with both ${\rm ^{87}Sr}$ and ${\rm ^{88}Sr}$.
While the bosonic isotope has a relatively simple level structure and a much higher natural abundance, a strong magnetic field of a few hundred Gauss is required to create a large enough transition linewidth~\cite{Taichenachev06} to produce the necessary atom-cavity interaction for superradiance~(Sect.~\ref{sec:las:model}).
On the other hand, the fermionic isotope will provide a much smaller flux of atoms into the cavity due to the reduced natural abundance, but the \transclock transition is already slightly allowed even at zero magnetic field through hyperfine interaction, with a linewidth of $2\pi \times 1.35$\,mHz~\cite{muniz_cavity-qed_2021}. Therefore, only a small magnetic field is needed to avoid degeneracy of energy levels and undesirable coherent effects.
For the reasons mentioned above, the following simulations concentrate mainly on the bosonic isotope.
The large bias magnetic field needed to open the clock transition in ${\rm ^{88}Sr}$ will be generated by the set of Helmholtz coils inside the shield, oriented along the $x$-axis. Here, the main purpose of the shielding is the magnetic isolation between the red MOT and the science chamber, which should allow us to maintain a continuous red MOT.
In this configuration, we can take full advantage of the large atomic flux of ${\rm ^{88}Sr}$ generated by our continuous system.

\section{Cooling and loading the atoms from the transport guide into the optical conveyor}
\label{sec:cool}
We will now focus on simulations of the atoms as they arrive in the science chamber. We first consider the three optical molasses beams, which are used to cool the atoms before and during transfer from the dipole guide and reservoir to the moving optical lattice (Fig.~\ref{fig:SC}b).
The first molasses beam (M1) addresses the ${\rm ^1S_0}\rightarrow{\rm ^3P_1}, m_J=0$ cooling transition and is used to slow the atoms in the guide beam that travel at high velocities before reaching the reservoir. The second beam (M2) is polarized in ($y$-$z$-plane) and acts on ${\rm ^1S_0}\rightarrow{{\rm ^3P_1}, m_J=1}$ and propagates parallel to the first molasses beam, but addresses the atoms in the region where the dipole guide, the reservoir, and the cavity mode overlap.
Finally, the third beam (M3), acting on ${\rm ^1S_0}\rightarrow{{\rm ^3P_1}, m_J=1}$, also interacts with atoms in the reservoir but propagates perpendicular to the dipole guide. 
The latter two beams cool the atoms to a mean energy low enough to be trapped and drawn in by the optical conveyor.
\subsection{Average energy of trapped atoms in the reservoir}
\label{sec:cool:temp}

We first simulate the average energy of the atoms once they are collected in the reservoir.
We discuss two key aspects of the cooling and loading process: the differential light shift on the cooling transition caused by the various light fields and the scattering dynamics of the cooling process.
We calculate the light shifts using the expression for the polarizability given in Appendix~\ref{App:a} and simulate the dynamics of the cooling process using a semiclassical Monte-Carlo method (SCMC), explained in detail in Appendix~\ref{App:B}. 
With the SCMC method, we calculate the probabilities of photon scattering from different molasses beams by individual atoms with specified positions and velocities, assuming their internal state is in local equilibrium.

The dipole potentials of the guide beam, reservoir, and optical lattice for the ${\rm ^1S_0}$, ${{\rm ^3P_1}, m_J=0}$ and ${^3P_1, m_J=\pm1}$ states will cause significant position-dependent light shifts on the cooling transitions addressed by the optical molasses beams.
First, let us consider only the effect of the differential light shift from the dipole guide and reservoir beams.
The dipole guide beam propagates along the $\hat{z}$-axis, so for this beam, the potential gradient only affects the $(x,y)$-plane, as the light shift is essentially constant along the $z$-axis. 
To slow down as many atoms in the dipole guide as possible, we can modulate the frequency of the optical molasses beams to address different velocity classes.
Instead of modulating frequency in time, we take into account multiple frequencies simultaneously.
We can choose the frequency range of the first molasses beam to be resonant with the atoms closest to the center of the dipole guide.
However, as the atoms approach the reservoir, the change in light shift along the $\hat{z}$ direction becomes significant, as the reservoir beam propagates $5^{\circ}$ from the $\hat{z}$-direction in the $(x,z)$-plane. 
This makes the capture velocity position dependent also along $\hat{z}$, as shown in Fig.~\ref{fig:trap}.
This dependence must be considered when choosing the frequencies of the second and third optical molasses beams, as they interact with atoms in the reservoir.
\begin{figure}
    \centering
\includegraphics[width=0.47\textwidth]{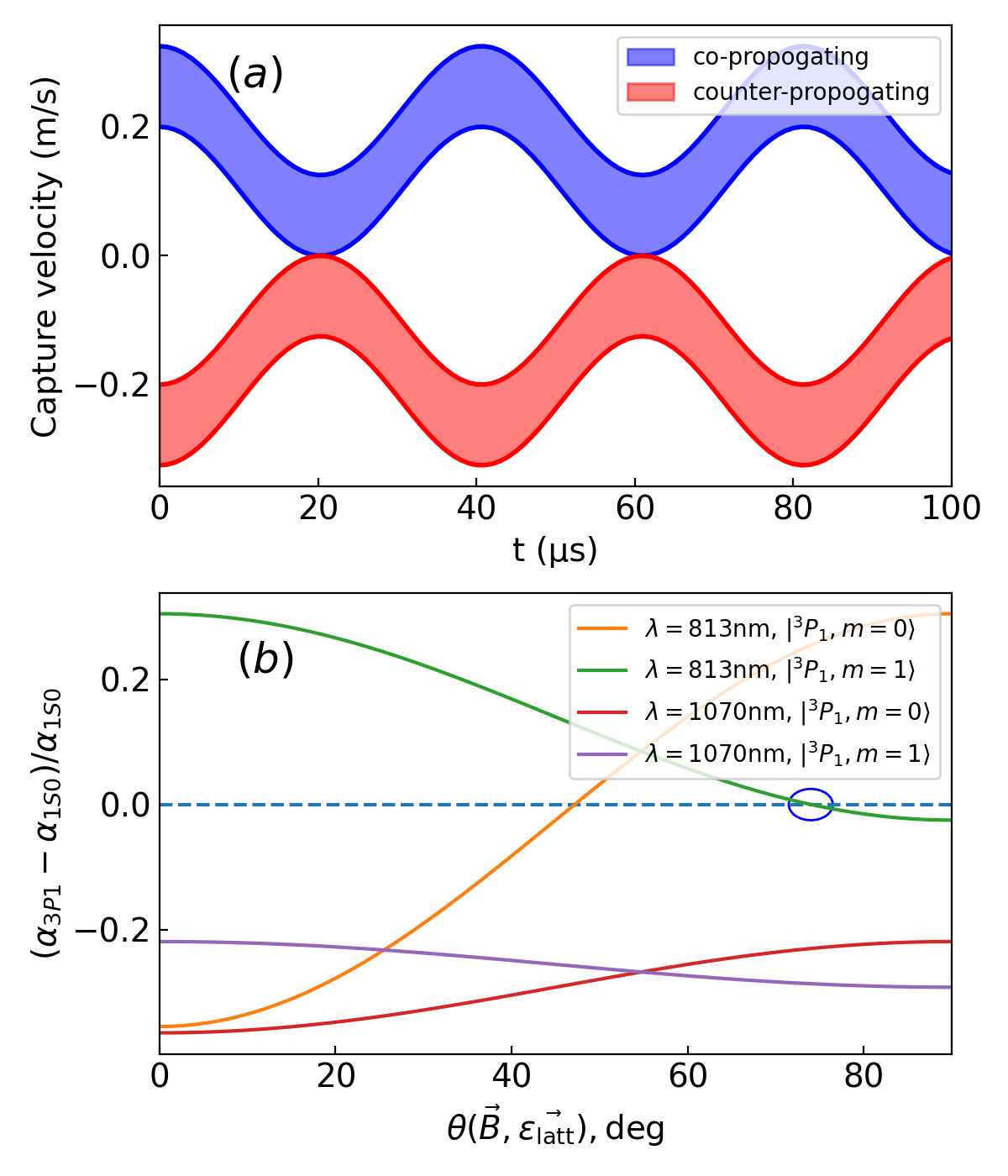}
    \caption{(a) The time-dependent oscillation of the capture velocity range as seen by atoms at a fixed position ($z=10\,\mu$m). This oscillation originates from the optical lattice moving with a velocity of $1\,$cm/s. (b) Polarizability of the atoms as a function of the angle between the bias magnetic field $\vec{B}$ and the optical lattice polarization. At an angle of $\approx 74^{\circ}$ between the optical lattice polarization and the B-field, the ${{\rm ^3P_1},m=0} \rightarrow {\rm ^1S_0}$ transition becomes magic due to the tuned tensor polarizability term.}
    \label{fig:Freq_O}
\end{figure}
By using a range of frequencies addressing different velocity classes, we find that for an average incoming velocity of 10~cm/s, we can reduce the average energy of atoms trapped in the reservoir to approximately $12\,\mu$K (Fig.~\ref{fig:trap}d). Each atom expects approximately 300 recoil photon events inside the reservoir.
\subsection{Loading of atoms into the optical lattice}
\label{sec:cool_load}
As the atoms approach the optical lattice, the situation becomes more complicated.
To simulate loading into the moving optical lattice, we must also consider the differential light shift caused by the substantial depth of the lattice, resulting in a significant change in the capture velocities we previously calculated. 
Because we are using a \textit{moving} optical lattice, the capture velocity range for the second and third optical molasses beams will oscillate with the motion of the optical lattice.
Consequently, scattering events will decrease as atoms move toward the optical lattice since the laser frequencies align only around the maximum value of the light shift (Fig.~\ref{fig:Freq_O}).

\begin{figure}[h!]
    \centering
    \includegraphics[width=0.47\textwidth]{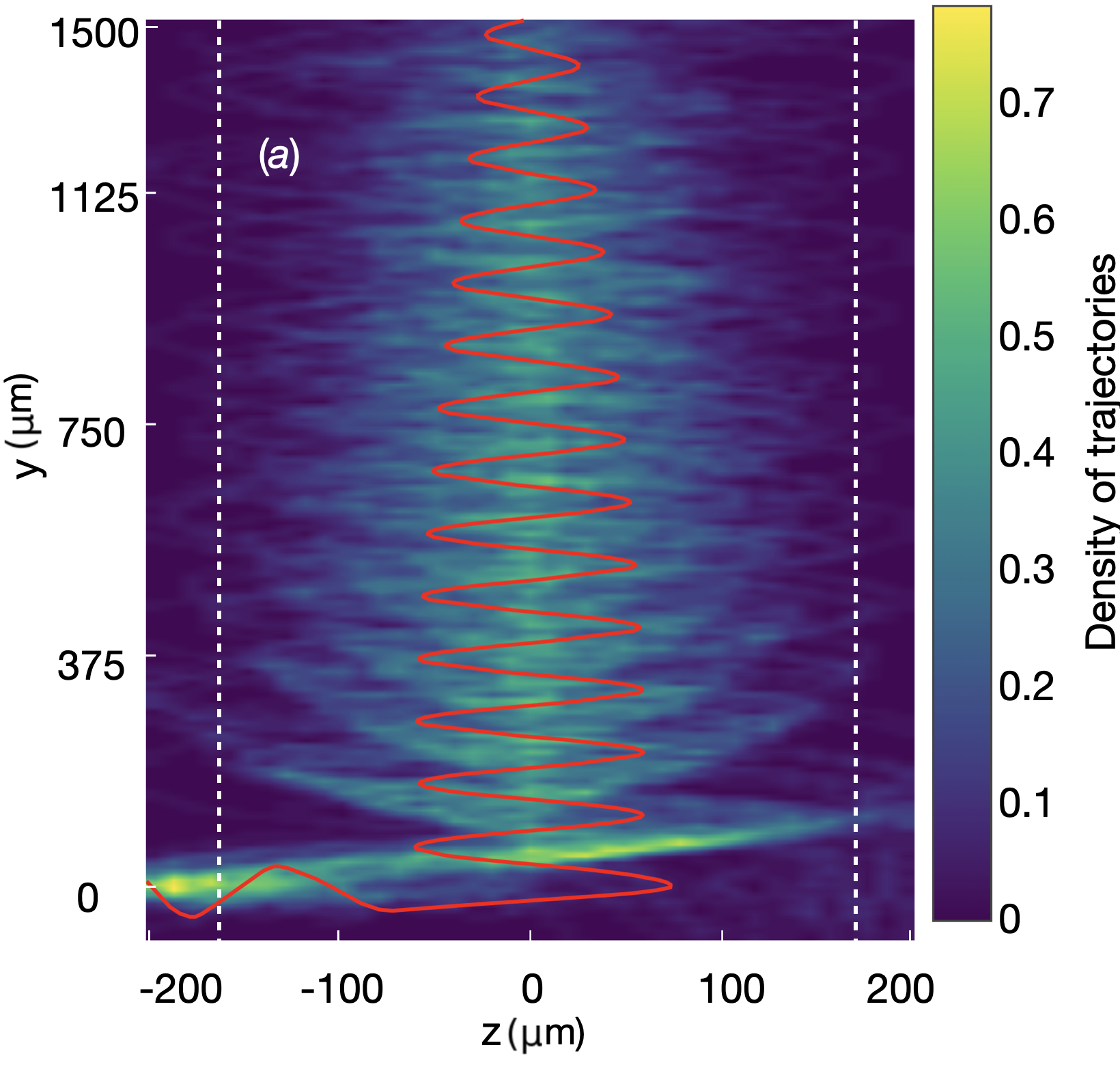}
    \includegraphics[width=0.47\textwidth]{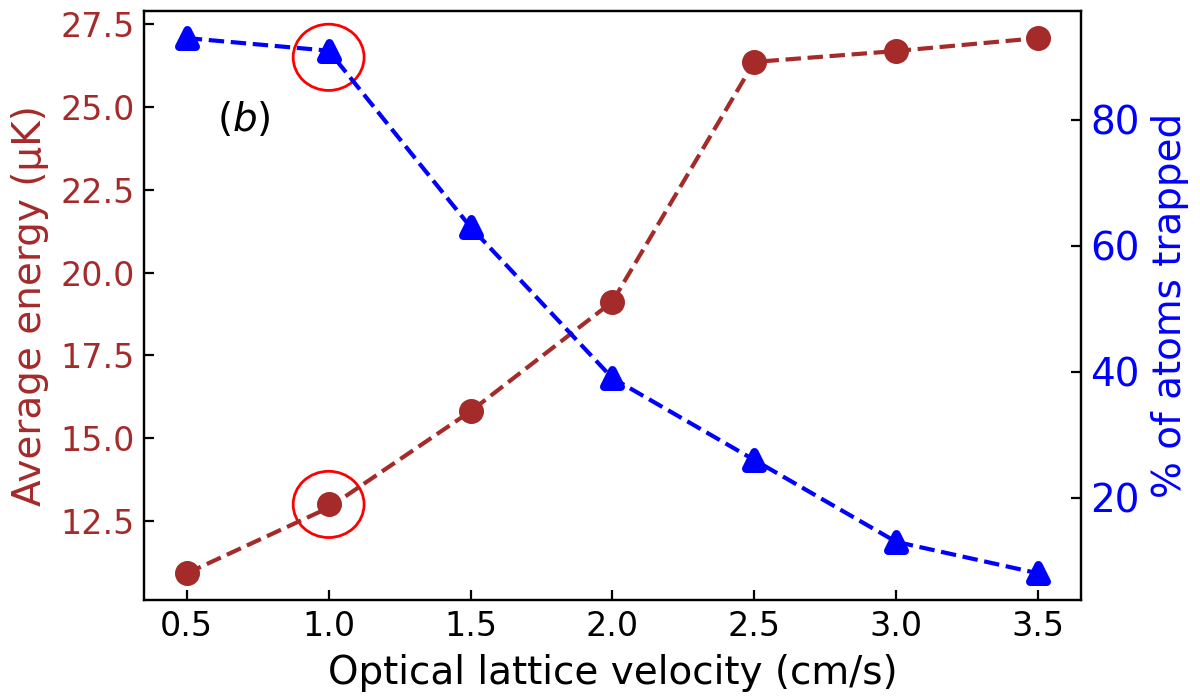}
    \caption{(a) Density of 100 trajectories of atoms as they travel along the moving optical lattice (velocity of the optical lattice = 1\,cm/s and potential depth = 30\,$\mu$K).
    A typical trajectory is shown by the red trace.
    The outline of the optical lattice is indicated by the white dashed lines. 
    (b) Average energy in $\mu$K (brown circles) and percentage of atoms trapped (blue triangles) as a function of the optical lattice velocity is presented.}
    \label{fig:Tra}
\end{figure}

Again using SCMC, but now including the time-dependent optical lattice light shift, the scattering rate is explicitly time-dependent (\ref{app:ss_ol}). 
Due to the oscillation of the capture velocity range caused by the moving lattice, there is an increase in the average energy distribution of the atoms, which reaches a value of approximately 20~$\mu$K. 
However, this increase in average energy can be minimized by tuning the Tensor polarizability term of the transition (${\rm ^1S_0}\rightarrow{\rm ^3P_1},\,m_J=1$) to a magic polarization (Fig.~\ref{fig:Freq_O}(b)).
We found that with a lattice speed of approximately 1\,cm/s, we achieve a $91\%$ loading efficiency into the lattice, with an average energy of $\rm 12.5\, \mu K$, yielding the simulated atomic trajectories in Fig. ~\ref{fig:Tra}(a).
There is, however, a tradeoff between the conveyor speed and the energy and number of atoms trapped.
If the conveyor velocity is too high, the average energy of the trapped cloud increases and the percentage of trapped atoms sharply drops off (Fig.~\ref{fig:Tra}(b)).
\section{Preparation of the atoms in the upper lasing state}
\label{sec:pump}
\begin{figure}
    \centering
    \includegraphics[width=0.43\textwidth]
    {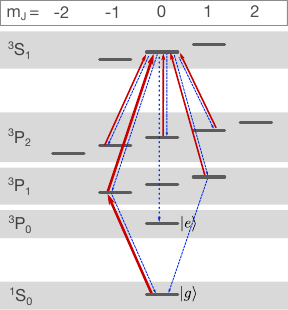}
    \caption{The strontium level structure relevant for our pumping scheme from ${\rm ^1S_0}$ to ${\rm ^3P_0}$ is shown here, including the Zeeman sublevels. The thicker red arrows indicate the main pumping transitions, while the thinner red arrows represent the relevant repumping transitions. The possible spontaneous decay channels are depicted as blue dashed arrows.}
    \label{fig:pump}
\end{figure}
After loading atoms into the optical lattice, we must pump them into the ${\rm ^3P_0}$ state. A possible pumping scheme compatible with our experiment is presented in Figure~\ref{fig:pump}. 
In the presence of the strong magnetic field required to open the clock transition in $^{88}$Sr, the Zeeman sublevels of the $^3$S$_1$, $^3$P$_2$, and $^3$P$_1$ states will split, allowing them to be independently addressed. 
To sufficiently populate ${\rm ^3P_0}$, we first pump atoms to the ${\rm ^3S_1}$ state.
From there, atoms can decay to ${\rm ^3P_0}$, but most likely will decay to the unwanted states, ${\rm ^3P_1},\,m_J=\pm1$ and ${\rm ^3P_2},\,m_J=-1,0,1$ (but not into ${\rm ^3P_1}, m_J = 0$, as this decay is prohibited by angular momentum selection rules). 
We must repump atoms out of these unwanted states, especially the ${\rm ^3P_2}$ states, as they are particularly long-lived~\cite{kluesener2024}. A more detailed description can be seen in Appendix~\ref{App:C}.
The pumping beam has a waist of $250\,\rm{\mu m}$ and is aligned along the z-axis, with its center positioned 2\,$\rm{mm}$ away from the reservoir along the y-axis. With $\vconv=1\, {\rm cm/s}$, we have an interrogation time of 5\,${\rm ms}$. To decrease heating, we pump from both directions. Using this scheme, $83\%$ of the atoms can be successfully loaded into the lattice and pumped to the $^3$P$_0$ state, while the rest are lost somewhere in the loading and pumping process. The fraction of loaded atoms remaining in the ground state after pumping is negligible. 

Atoms will undergo an average of 12 photon recoils throughout the pumping process, leading to an average energy of the pumped sample around $16\,\mu$K. 
We choose an optical lattice depth of $30\,\mu$K, which will keep the atoms in the optical lattice long enough so that they can contribute to superradiant emission.
Elastic collisions, which could create higher energy atoms, are negligible under our conditions.
\section{Simulation of the superradiant laser output}
\label{sec:las}
To determine the parameters necessary to achieve continuous superradiant lasing, we have numerically simulated the intra-cavity field for atoms traveling in the optical conveyor along a running-wave cavity mode resonant with the \transclock transition. We use a semiclassical model to analyze the impact of systematic effects, particularly from density shifts, light shifts, decoherences and losses, and the inhomogeneity of external magnetic fields. In our simulations, we assume a total roundtrip length $\lcav=20\,{\rm cm}$ of the bowtie cavity and a cavity finesse $F=5\times 10^4$. This corresponds to a cavity field energy decay rate equal to $\kappa = 2\pi \times 150{\,\rm kHz}$. We also take the speed of the optical conveyor as $v_{\rm conv}=1{\,\rm cm/s}$ and the travel distance of the atoms before being ejected as $\lconv=2~{\rm cm}$.

To find the optimal atomic flux, we begin with a simplified model and consider two-level atoms that are loaded into the optical conveyor in the upper lasing state. 
The atoms are then carried along the conveyor for a time $\tau$ before being ejected from the cavity by the push beam. 
We suppose that one of the running-wave cavity modes is resonant with the atoms in the lattice, taking into account the first-order Doppler shift.
We also assume that the interaction time $\tau$ of the atom with the cavity field is much shorter than all the inverted relaxation rates of the atomic degrees of freedom. 
The Hamiltonian describing such a model in the respective rotating frame has the form
\begin{equation}
\HH = \hbar 
g \sum_j \Gamma^j(t)\left[\a \sig^j_{eg}+\ad \sig^j_{ge} \right],
\label{eq:conv:1} 
\end{equation}
where $g$ is the coupling strength between the atomic transition and the cavity field, $\a$ and $\ad$ are the annihilation and creation operators of the cavity mode, and $\sig^j_{eg}=\ket{e_j}\bra{g_j}$ and $\sig^j_{ge}$ are the rising and lowering operators of the $j$th atom. Here, we have also introduced functions $\Gamma^j(t)=\Theta(t-t_j)-\Theta(t-t_j-\tau)$, which describe the time-dependence of the atom-cavity coupling, where $t_j$ is the time of injection of the $j$th atom into the conveyor. With coarse-grained time averaging, $\sum_j \delta(t-t_j) \approx \sum_j \delta(t-t_j-\tau) \approx \Phi$, where $\Phi$ is the atomic flux. In this scenario, a stationary solution for the intracavity field can be found from the equation~\cite{Yu08} 
\begin{equation}
\sin^2 \chi = \chi^2 A^2, \quad {\rm where} \quad A = \sqrt{\frac{ \kappa}{\Phi g^2 \tau^2}}.
\label{eq:conv:2} 
\end{equation}
Here $\chi=g a \tau$, $a=\langle \a \rangle$ is the cavity field in the mean-field approximation, and $\kappa$ is the decay rate of the energy of the cavity mode. This equation has a single solution, assuming the parameter $A$ lies inside the following interval:
\begin{equation}
0.21723...<A < 1.
\label{eq:conv:3} 
\end{equation}
If $A>1$, Eq.\,(\ref{eq:conv:2}) has no non-zero solutions, which corresponds to no superradiant emission, and if $A<0.22$, Eq.\,(\ref{eq:conv:2}) has multiple non-zero solutions, which can lead to unstable superradiant emission. Note that for real systems with non-negligible dephasing of the lasing transitions, stable solutions can also exist for $A<0.22$.

In the following subsections, we consider two different semiclassical models of an optical conveyor laser carrying ${^{88}}$Sr atoms, which include collision-induced effects, as well as position-dependent shifts caused by inhomogeneity of the magnetic field. 
The first ``basic'' model is based primarily on the data reported in Ref.~\cite{Lisdat09}, whereas the second one considers a more extensive estimation of the collision-induced dephasing. 
For both cases, we simulate the intracavity field $\langle \hat{a}\rangle $ and study the dependence of the number of intracavity photons $n=\langle \ad \a \rangle$ and frequency shift $\Delta_{\rm out}$ on the atomic flux $\Phi$. We also investigate the influence of the position-dependent frequency shift created along the conveyor due to magnetic field inhomogeneities.
The number of intracavity photons $n$ can be used to calculate the output power $P_{\rm out}$ of the superradiant laser as
$P_{\rm out}= \hbar \omega n \kappa \eta$, where $\eta$ is the probability for the photon to leave the cavity through the outcoupling mirror.
\subsection{${\rm ^{88}Sr}$ atoms in an optical conveyor lattice in the presence of a magnetic field: the basic model}
\label{sec:las:model}
\begin{figure}
    \centering
    \includegraphics[width=0.47\textwidth]{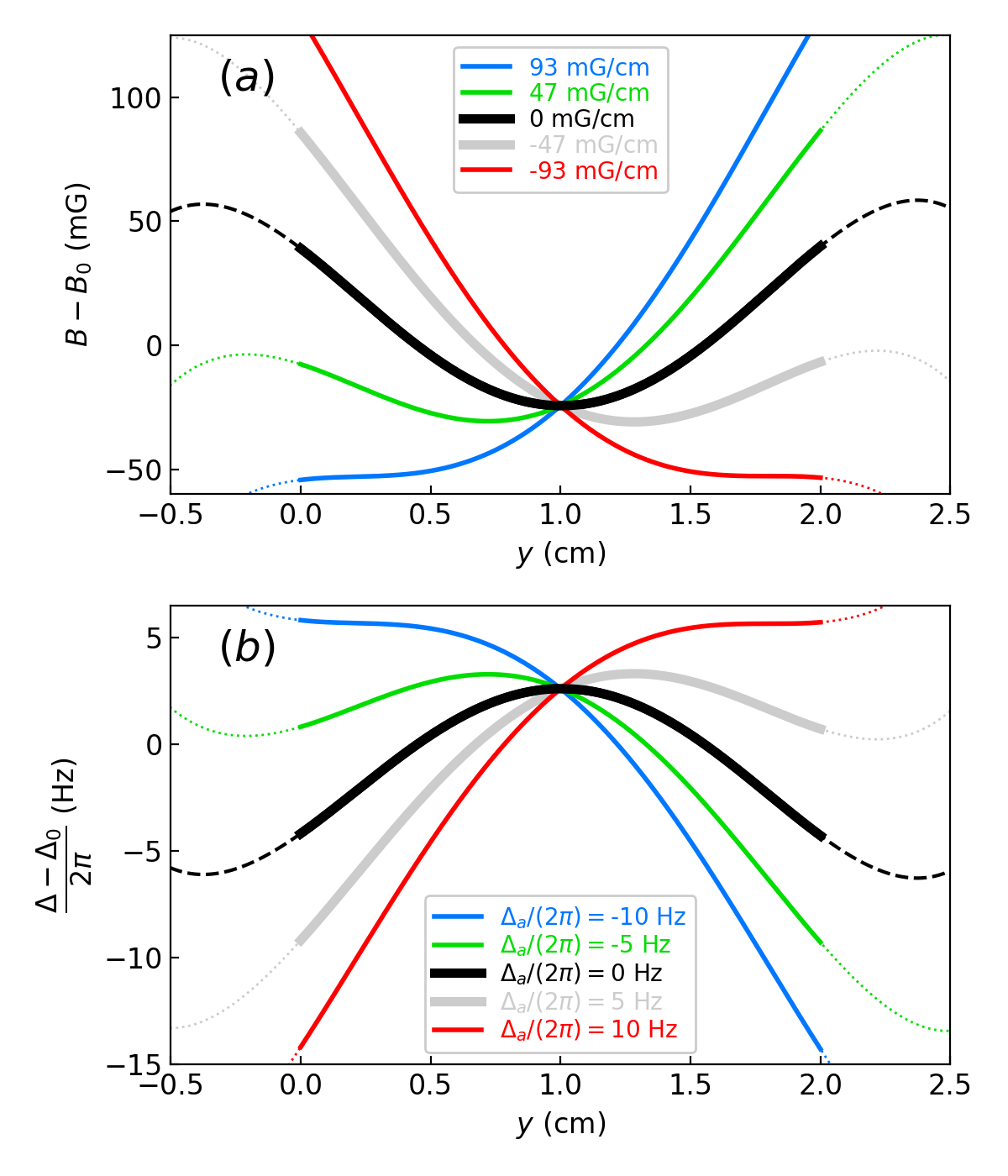}
    \caption{Spatial inhomogeneity of magnetic field and the resulting clock transition frequency change. (a) Magnetic field magnitude $B$ along the direction of the optical conveyor and (b) corresponding second-order Zeeman shift, referenced to an offset field of $B_0=230~{\rm G}$, which corresponds to a frequency shift of $\Delta_0\approx -2\pi\times12.3~{\rm kHz}$ as given by Equation~\ref{eq:conv:4}. The segments that are rendered solid are within the emission zone. The black curve corresponds to the simulated magnetic field of the Helmholtz coil pair. The colored curves represent the presence of an additional magnetic field gradient. The resulting position-dependent frequency shift is shown in (b) with the same color encoding. }
    \label{fig:mf}
\end{figure}

The single-photon \transclock transition in neutral bosonic Sr is forbidden to all orders of multipole expansion. However, this transition can be partially allowed in the presence of an external field. For example, a static magnetic field $\vec{B}$ \cite{Taichenachev06}, can slightly mix the $|{\rm ^3P_0}\rangle$ and $|{\rm ^3P_1},m_J=0\rangle$ states, opening the \transclock transition.
This comes at the expense of a change in the transition frequency due to the second-order Zeeman shift $\Delta_{\rm mg}=\delta \omega_{\rm {^3P_0}\rightarrow{^1S_0}}$.
The induced E1 transition rate $\gamma_{\rm {^3P_0}\rightarrow {^1S_0}}$ can be calculated from the second-order Zeeman shift as:
\begin{align}
\Delta_{\rm mg} = \gamma_{\rm {^3P_0}\rightarrow {^1S_0}} \frac{\omega_{\rm ^3P_1\rightarrow ^3P_0}}{\gamma_{\rm {^3P_1}\rightarrow {^1S_0}}}=\beta B^2,
\label{eq:conv:4}
\end{align}
where $\beta \approx -2\pi \times 23.3~{\rm MHz/T^2}=-2\pi \times 233~{\rm mHz/G^2}\approx -1.464~{\rm s^{-1}/G^2}$ \cite{Taichenachev06}. From this expression, we can clearly see the high sensitivity of the clock transition frequency to variations in the applied bias magnetic field as the frequency shift scales with $B$.
In our case, the atoms will be most sensitive to spatial inhomogeneities and fluctuations of the magnetic field within the emission zone of the bowtie cavity.

In order to open the \transclock transition, we apply a fairly strong and homogeneous bias magnetic field $B \approx B_{0}=230~{\rm G}$.
This field strength corresponds to a \transclock transition linewidth of $\gamma_{\rm {^3P_0}\rightarrow{^1S_0}} \approx 2\pi \times 16.4~{\rm \mu Hz}$ and a second order differential Zeeman shift of $\Delta_{\rm mg} \approx \Delta_{0}- 2\pi \times 107~{\rm mHz/mG}\times(B-B_0)$, where $\Delta_{0}=\beta B_0^2 \approx -2\pi \times 12.3~{\rm kHz}$.
To evaluate the dependence of the second-order Zeeman shift $\Delta_{\rm mg}$ on the position along the conveyor, we first determined the distribution of the magnetic field strength $B$ with COMSOL, the results of which are shown by the black curve in Figure~\ref{fig:mf}. 
The simulated magnetic field inhomogeneity in the center of the emission zone is mostly due to the incompletely closed magnetic shielding, where holes are required for optical access and connection to the rest of the vacuum system. 
The simulated imperfections lead to a position-dependent frequency shift $\Delta_{\rm mg}(y)$ that becomes more significant as the field strength increases. To counteract this problem, we consider the possibility of adding an extra gradient $G_B$ to the bias field $B$, which will add an additional position-dependent shift $\Delta_a (2y/\lconv-1)$. Therefore, the overall position-dependent shift caused by the magnetic field has the form
\begin{align}
\Delta(y)=\Delta_{\rm mg}(y)+\Delta_a (2y/\lconv-1).
\label{eq:conv:5}
\end{align}
In this expression, the amplitude $\Delta_a$ of the extra position-dependent shift can be calculated from the magnetic field gradient $G_B$ as $\Delta_a=\beta B_0 G_B \lconv$.
This position-dependent shift can help us partially compensate irregularities in $\Delta_{\rm mg}$ in parts of the emission region. Interestingly, it also allows us to compensate for the collisional shifts (\ref{eq:conv:11}), which depend on the densities of the atoms in the ground and the excited states. The total collisional shift varies as the atoms move along the conveyor, as shown in Figure~\ref{fig:10}(g).

In our analysis, we use mean-field equations where we suppose that each atom interacts with the cavity field created by the atoms themselves. Quantum correlations between different atoms have been neglected. 
We also take into account collisional decoherence, shifts, and losses, which have been adapted from Ref.~\cite{Lisdat09}. 
We suppose that the atoms interact only with the self-generated, running-wave cavity mode which co-propagates with the optical conveyor, as this cavity mode is resonant with the atomic transition, while the counter-propagating running-wave mode, present in every ring cavity, will be detuned by about $\delta\omega_{\rm Doppler}=2 \omega \vconv/c \approx  2\pi\times 28.6~{\rm kHz}$ at $\vconv=1~$cm/s. 
This detuning suppresses lasing on the counter-propagating mode and hinders atoms from collectively interacting with this mode due to the mismatch of the relative phases.

To reduce the computational cost, we group the atoms into $M$ clusters distributed along the optical conveyor, with all the atoms of the same cluster having the same internal states. Each cluster occupies a segment of length $\lc = \lconv/(M-1)$ centered at position $y_j$ along the conveyor. The clusters are initialized at position $y_{j,0}\,=\,-\lc/2$, and the atoms get removed only when they reach the position $y_{j,f}=\lconv+\lc/2$, which corresponds to the end of the emission zone. When $-\lc/2<y_j<\lc/2$ or $\lconv-\lc/2<y_j<\lconv+\lc/2$, the coupling coefficient $g$ between the atoms and the cavity field is multiplied by the fraction of atoms in the cluster inside the emission zone of the conveyor.  The number $N^j$ of atoms in the $j$th cluster is randomly distributed around $\Phi \lc/\vconv$, where we have used a Poissonian distribution.

The mean-field equations for the cavity field $a=\langle \a \rangle$ are
\begin{align}
\frac{da}{dt}&=-\left[ \frac{\kappa}{2}+i\delta_a \right]a - i \sum_j g(y_j) \sigma^j_{ge} N^j,
\label{eq:conv:6}
\end{align}
where $\delta_{a} = \delta_c - k_{0}/\vconv$, $\delta_c$ is the detuning of the cavity field from the rotating frame in which we consider the system, $\vconv$ is the speed of conveyor, and $k_0$ is the wave number of the cavity mode. The sum is taken over all atoms in the optical conveyor. As long as the cavity decay rate $\kappa$ is much larger than any shifts, decay rates, or decoherence rates, the field $a$ quickly equilibrates with the atomic degrees of freedom and can be adiabatically eliminated:
\begin{align}
a&=\frac{-2 i}{\kappa + 2 i \delta_a} \sum_j g(y_j) \sigma^j_{ge} N^j.
\label{eq:conv:7}
\end{align}
Next, we can adapt the equations for atomic coherences $\sigma^j_{ab}=\langle \sig^j_{ab}\rangle$ (where $\sig^j_{ab}=|a^j\rangle \langle b^j| $) of individual atoms from Ref.~\cite{Lisdat09} as follows:
\begin{widetext}
\begin{align}
\frac{d\sigma^j_{ge}}{dt}&=-\left[ \frac{\gamma+\gamma_e+\gamma_g+w(y_j)}{2}+\gamma_R+\nu_p(y_j) + \Gamma^j_{\rm coll} 
+ i\left(\Delta(y_j) +\Delta^j_{\rm coll} + \delta_{\rm p}(y_j) \right) \right] \sigma^j_{ge} 
+ ig(y_j) a (\sigma^j_{ee}-\sigma^j_{gg}),
\label{eq:conv:8} \\
\frac{d\sigma^j_{ee}}{dt}&=i g(y_j) \left[a^*  \sigma^j_{ge}-a \sigma^j_{eg} \right] 
-(\gamma_e+\gamma)\sigma^j_{ee} - n_j 
(\gamma_{ee}\sigma^j_{ee}+\gamma_{eg}\sigma^j_{gg})\sigma^j_{ee}
+ w (y_j) \sigma^j_{gg},
\label{eq:conv:9} \\
\frac{d\sigma^j_{gg}}{dt}&=- i g(y_j) \left[a^* \sigma^j_{ge}-a\sigma^j_{eg} \right]+ \gamma \sigma^j_{gg} 
-\left[\gamma_g+w(y_j)+n_j \gamma_{ge} \sigma^j_{ee}\right]\sigma^j_{gg},
\label{eq:conv:10}
\end{align}
\end{widetext}
where $\gamma=7.8\times 10^{-5}$ is the spontaneous transition rate at $B=230\,{\rm G}$, and $\gamma_{e}$ and $\gamma_g$ are density-independent inverse lattice lifetimes for the ground and the excited states. Here, we take $\gamma_{e}=\gamma_{g}=0.33~{\rm s^{-1}}$ as a conservative estimation, which corresponds to 3\,s of lattice lifetime. 
The position-dependent pumping rate is denoted by $w(y)$, and the shift and dephasing rates in the pumping zone are written as $\delta_{\rm p}(y)$ and $\nu_p(y)$, respectively. The extra density-independent dephasing rate caused by elastic collisions with a background gas and Raman scattering of photons from the optical lattice potential is denoted by $\gamma_R$~\cite{Doerscher18}. We have taken $\gamma_R=0.3~{\rm s^{-1}}$ and define the total rate of collision decoherence as
\begin{align}
\Gamma^j_{\rm coll}&=n_j \left[\frac{\sigma^j_{ee} \gamma_{ee}+ [\sigma^j_{gg}+\sigma^j_{ee}] \gamma_{ge}}{2} + \gamdep \sigma^j_{gg}\right]
\label{eq:conv:11}
\end{align}
and the collision shift as
\begin{align}
\Delta^j_{\rm coll}&=n_j [ \mu (\sigma^j_{ee}+\sigma^j_{gg})+\epsilon (\sigma^j_{ee}-\sigma^j_{gg}) ].
\label{eq:conv:12}
\end{align}
Here, we define the loss, dephasing, and shift coefficients as follows: $\gamma_{ee} = (4\pm 2.5) \times 10^{-12}~{\rm cm^3/s}$, $\gamma_{ge}  = (5.3\pm 1.9) \times 10^{-13}~{\rm cm^3/s} $, $\gamma_{\rm dep}  = (3.2\pm 1.0) \times 10^{-10}~{\rm cm^3 / s} $, $\mu = 2 \pi \times 8.2 \cdot 10^{-11}~{\rm cm^3 \cdot Hz}$, and $\epsilon = 0.33\mu$~\cite{Lisdat09}.
Details of the calculation of the number density $n_j$, coupling strength $g$, and other relevant parameters are given in Appendix~\ref{App:D}.

\subsection{Results of simulation for the basic model}
\label{sec:las:results}
\begin{figure*}
    \centering
    \includegraphics[width=1\textwidth]{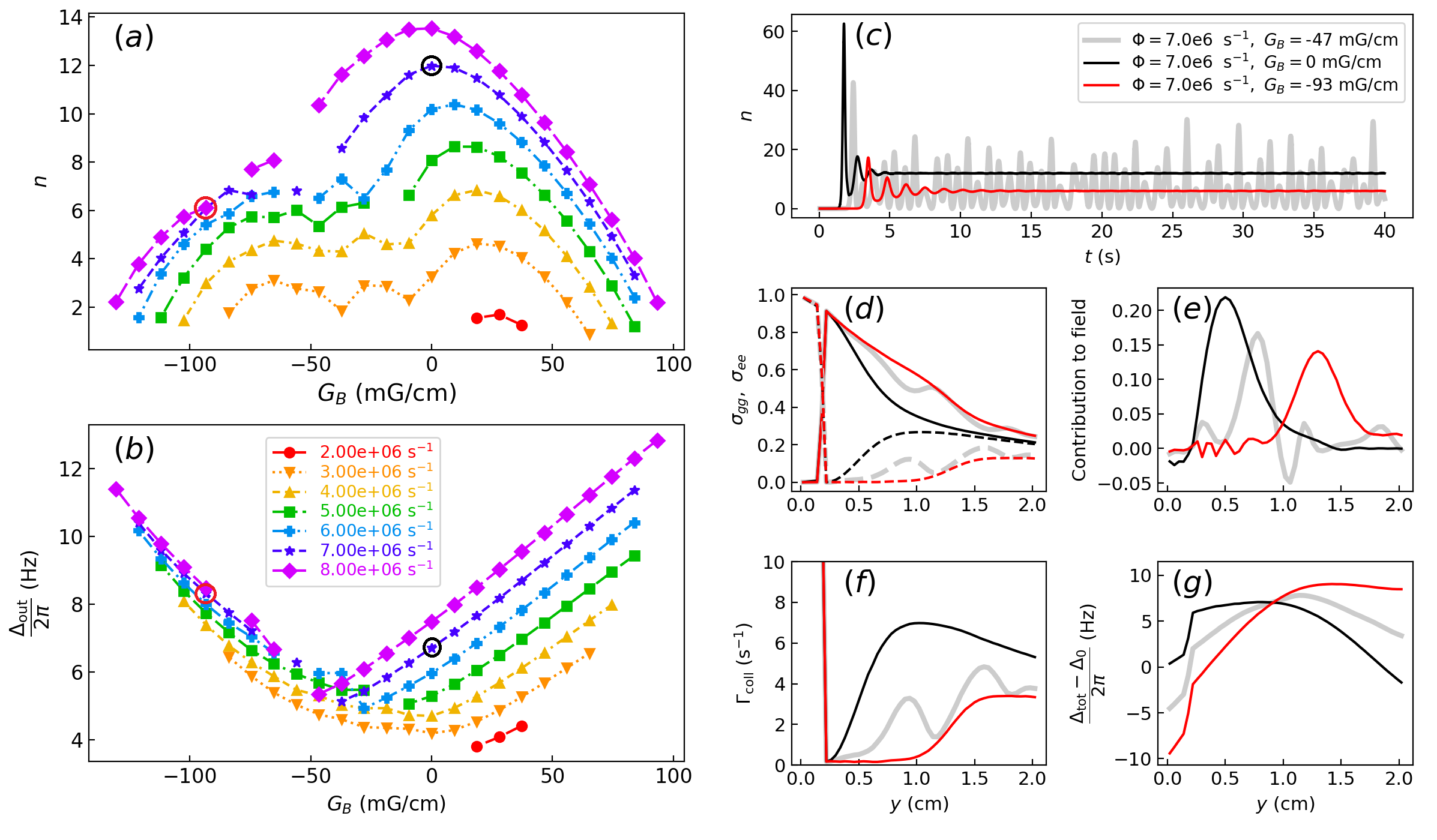}
   \caption{Results of superradiant lasing simulation with the basic model. Left: (a) simulated intracavity photon numbers and (b) frequency shifts of the output field relative to the atomic transition versus gradient $G_B$ of the magnetic field for different atomic fluxes $\Phi$ (see legend; same color code for (a) and (b)).  Here we use $B_0=230\,{\rm G}$ and $\FF=~50000$. Circled points correspond to stable solutions investigated in detail in right pane. 
   Right: (c) Examples of intracavity photon number over time for $\Phi=7\times 10^6~{\rm s^{-1}}$ and three different values of the magnetic field gradient $G_B$. (d) Position dependent distributions of populations $\sigma_{ee}$ (solid) and $\sigma_{gg}$ (dashed) along the conveyor. (e) Position-dependent contribution of the atoms to the cavity field $a$ is defined as the imaginary part of $\sigma_{ge} \exp[-i{\rm arg}(a)]$. (f) Distribution of collision-induced dephasing. (g) Total (magnetic plus collision-induced) shift $\Delta_{\rm tot}=\Delta(y)+\Delta_{\rm coll}$ along the conveyor. The color code for (c) -- (g) is shown in (c).}
    \label{fig:10}
\end{figure*}
We now present the results of numerical simulations of the superradiant laser output 
using the semiclassical model described above and with the collisional dephasing rate given by Eq.~(\ref{eq:conv:11}). 
We assume a total roundtrip
length $\lcav=20~{\rm cm}$ and cavity finesse $\FF=5\times 10^4$, which gives the decay rate of the  cavity field energy:
\begin{equation}
\kappa=\frac{2\pi c}{\FF \lcav}\approx 1.88 \times 10^5~{\rm s^{-1}}.
\label{eq:conv:16}
\end{equation}
The output laser power per single intracavity photon can be estimated as $\kappa \hbar \omega \eta\approx 1.34 \times 10^{-14}~{\rm W}$, where $\omega\approx 2\pi\times 429~{\rm THz}$ is the frequency of the \transclock transition and $\eta$ is the fraction of output power emitted through the outcoupling mirror. We assume all four mirrors have equal transparency, which leads to $\eta=0.25$.
As previously defined in Section~\ref{sec:cool}, the waist of the 813-nm magic wavelength optical lattice mode is $140~{\rm \mu m}$, and we take the waist $W_0$ of the resonant co-propagating 698-nm cavity mode as $W_0=130~\mathrm{\mu m}$.
We also make a conservative estimation of the temperature of the atomic ensemble, $T=10~{\rm \mu K}$, and define the depth of the optical lattice as $\Uconv=30~{\rm \mu K}$.

The atoms are loaded into the conveyor in the $^1$S$_0$ state and get pumped into the upper lasing state, as described in Section~\ref{sec:pump}. The pumping process is simulated using the position-dependent incoherent pumping rate $w(y)=w_0 p(y)$, the pumping-related dephasing rate $\nu_p(y)=\nu_p^0 p(y)$, and light shift $\delta_p(y)=\delta_p^0 p(y)$, where 
\begin{equation}
p(y)=\exp \left(-\frac{2(y-y_p)^2}{W_p^2} \right),
\label{eq:conv:22}
\end{equation}
$y_p=2~{\rm mm}$, and $W_{p}=250~{\rm \mu m}$. We take $w_0=270~{\rm s^{-1}}$ and $\nu_p^0=400~{\rm s^{-1}}$ as a typical values~(Appendix~\ref{App:C}). 
At this point, we set the pumping-induced light shift, $\delta_p^0=0$. Later, however, we will show that reasonable values of pumping-induced light shift will have only a minor influence on the amplitude and frequency of the output laser field.
Using these parameters, we perform a series of simulations of the superradiant laser output for different atomic fluxes $\Phi$. 

Note that the mean-field equations (\ref{eq:conv:6}) and (\ref{eq:conv:8}) - (\ref{eq:conv:10}) are invariant to a common phase shift of atomic coherences $\sigma^j_{ge}$ and cavity field $a$. To break this phase symmetry and initiate the lasing process, we assume that, at the beginning of the simulation, the atomic ensembles in the cavity have some small ``seed'' populations and coherence: $\sigma^j_{ee}=(1-\cos(\theta_0))/2$, $\sigma^j_{gg}=(1+\cos(\theta_0))/2$  $\sigma^j_{ge}=\sin(\theta_0) \exp(i \phi_0^j)$, where $\theta_0=0.07~{\rm rad}$ and $\phi_0^j$ are randomly distributed between 0 and $2\pi$. All the atomic ensembles loaded into the conveyor after that have no ``seed'' coherence.

We first consider the results of simulations of the intracavity photon number $n$ in steady-state and the shift $\Delta_{\rm out}$ of the output radiation frequency with respect to the \transclock transition as a function of atomic flux $\Phi$ and magnetic field gradient $G_B$, shown in Fig.~\ref{fig:10}(a,b).
We set $B_0=230~{\rm G}$ and the number of clusters $M=51$. 
We choose $\Phi$ in the range $2.4\times 10^5~{\rm s^{-1}}\leq \Phi\leq 8\times 10^6~{\rm s^{-1}}$, which corresponds to $1\geq A \geq 0.173$.
The full simulated time is 40~s, but we truncate the first 20~s and use the last 20~s to calculate the characteristics of the signal once stabilized. Only stable solutions, where the variations of the amplitude of the intracavity field over the last half of the simulation period were less than $10~\%$ of the mean, are presented in Fig.~\ref{fig:10}. We can see that for some combinations of $(\Phi, G_B)$, the solutions are unstable. Notably, we see no laser output for $\Phi<2\times 10^6\,{\mathrm s^{-1}}$ , which corresponds to $A>0.346$, indicating a threshold atomic flux for superradiant emission.

To further investigate the lasing process, we consider simulations for $\Phi=7\times 10^6~{\rm s^{-1}}$ at three different values of the magnetic field gradient: $G_B=-47~{\rm mG/cm}$, $-93~{\rm mG/cm}$, and $0~{\rm mG/cm}$. Simulated time-dependent intracavity photon numbers for these three cases are presented in Figure~\ref{fig:10}(c), and one can see that for $G_B=0~{\rm mG/cm}$ and $-93~{\rm mG/cm}$, the solution is stable, whereas for $G_B=-47~{\rm mG/cm}$, we have an unstable intracavity field with irregular superradiant pulses. 

In Figure~\ref{fig:10}(d), we show the position dependence in the emission zone for the populations of the ground ($\sigma^j_{gg}$) and excited ($\sigma^j_{ee}$) states at the end of the simulation ($t=40~{\rm s}$) for these same parameters. 
The decrease of $\sigma_{ee}+\sigma_{gg}$  along the length of the emission zone corresponds to loss of atoms from the conveyor.
For $G_B=0~{\rm mG/cm}$ (black curves), the atoms return to the ground state faster than for $G_B=-47~{\rm mG/cm}$ and $G_B=-93~{\rm mG/cm}$ (gray and red curves, respectively).
This is because for $G_B=0~{\rm mG/cm}$, the atoms are coherently coupled with the cavity field primarily in the first half of the optical conveyor, whereas for $G_B=-93~{\rm mG/cm}$, they are coupled in the second half, which is in agreement with the atomic contribution to the intracavity field plotted in Fig.~\ref{fig:10}~(e). For $G_B=-47~{\rm mG/cm}$ (unstable regime), we see population oscillations between the ground and the excited state and similar oscillations in the intracavity field. 
We note that at the end of the emission zone, $\sigma_{ee}$ is still larger than $\sigma_{gg}$ for all three curves, indicating that less than half of the energy stored in the \transclock transition gets converted into the energy of the cavity field. 

We define the contribution of the single-atom coherence to the intracavity field as ${\rm Im}(\sigma_{ge} \exp(-i {\rm arg}(a))$, which is consistent with Eq.~\ref{eq:conv:7}.
If we look more closely at the position-dependence of this quantity in Fig.~\ref{fig:10}(e), we see that it is consistent with where the variation in the overall (magnetic plus collision) shift is minimal.
For example, for $G_B=0~{\rm mG/cm}$, the main contribution to the intracavity field is given by the atoms in the first half of the emission zone, and this corresponds to a plateau in the overall shift plotted in Fig.~\ref{fig:10}(g). 
Similarly, for $G_B=-93~{\rm mG/cm}$, the main contribution comes from the last half of the emission zone, but because a significant fraction of the atoms have been lost from the conveyor at this point, the amplitude of the signal is smaller.
For the unstable regime where $G_B=-47~{\rm mG/cm}$, we see random absorption and emission events between the atoms and the cavity field. These oscillations of energy lead to chaotic behavior of the out-coupled laser field.

In Figure~\ref{fig:10}~(f,g), we present position-dependent collisional dephasing rates $\Gamma_{\rm coll}$ and the total (magnetic and collision-induced) frequency shifts over the length of the optical conveyor. For all three values of $G_B$, the collisional dephasing rate is the highest at the beginning of the emission zone, when the atoms are still in the ground state, but abruptly decreases at $y=0.2~{\rm cm}$, when the atoms get pumped to ${\rm ^3P_0}$. 
The dephasing rate then starts to grow as atoms emit and are transferred from the excited state back to the ground state.
This is consistent with our model, in which the dominating source of dephasing is ground state collisions~(\ref{eq:conv:11}).
Eventually, dephasing decreases again as the total number density decreases due to losses. 

Finally, we can compare the plots of the total position-dependent shift in Fig.~\ref{fig:10}(g) and the magnetic shift presented in Fig.~\ref{fig:mf}(b) for the same B-field gradient to determine the effect of the collision shift on the total shift.
The variation of the collision shift throughout the emission zone is a dynamic process, as it will decrease as atoms are lost from the optical conveyor but increase as atoms decay from the excited state back down to the ground state.
Qualitative comparison of the two plots suggests that the magnetic shift dominates throughout the emission zone, and this variation does not have a significant effect on the total shift.

We can now estimate the fraction of atoms contributing to the intracavity field and how much output lasing power we can expect in such an experiment.
For a flux of $7\times10^6~$s$^{-1}$, the maximum power that can, in principle, be emitted into the cavity mode is $\hbar \omega \Phi \approx 2~$pW.
We calculate the power actually emitted into the cavity field as $P_{\rm field}=\hbar \omega \kappa n$.
For $G_B=0$ and $n \approx 11.9$, we get $P_{\rm field}=0.64~$pW.
Therefore, about 32\% of the atoms that are pumped to the excited state contribute to the intracavity field. This can be explained by the loss of excited state atoms from the optical conveyor and by atoms making only partial transfers between the excited and ground states.
As a result, the output lasing power on the clock transition that we can expect to achieve under optimal conditions is $P_{\rm out}=\eta P_{\rm field}=0.16~$pW, where we have again assumed $\eta=0.25$.

To investigate the sensitivity of the laser output to the pumping-induced light shift $\delta_p$, we perform additional simulations for $\delta_p^0=2\pi \times 25~{\rm Hz}$, a typical order-of-magnitude estimate of the effective light shift, and $\delta_p^0=2\pi \times 500~{\rm Hz}$, a more pessimistic upper bound, for the marked points in Fig.~\ref{fig:10}(a,b) (Table~\ref{table:EquivParameters}).
We find that the stability of solutions with the same $\Phi$ and $G_B$ does not depend on $\delta_p^0$, and for stable regimes, the difference in amplitude of the intracavity field for solutions with the same values of $\Phi$ and $G_B$, but different $\delta_p^0$, differs by less than a few percent. 
In addition, the frequency shift of the output field changes by less than $50~{\rm mHz}$ with change of $\delta_p^0$ from 0 to $500~{\rm Hz}$, which lies on the edge of the Fourier-limited resolution of our simulation. 
Such robustness can be explained by the fact that the atoms do not contribute to the intracavity field while they are affected by the pumping-induced light shift $\delta_p$, as the pumping zone is about 20 times smaller than the emission zone.
The huge dephasing associated with this pumping also helps to reduce its influence on the output of the superradiant laser.

The quantum noise-limited linewidth of the superradiant output can be estimated using the second-order cumulant expansion~\cite{Bychek2021, Kazakov22}. This method is based on clustering the atomic ensemble according to positions in the optical conveyor and considering the collision-induced shifts and loss and dephasing rates as external parameters, pre-calculated with the help of the semiclassical model described above. 
For $\Phi=7\times 10^6$ atoms/s the quantum noise-limited linewidths are about $5~{\rm \mu Hz}$ for $G_B=0$, and $7~{\rm \mu Hz}$ for $G_B=-93~{\rm mG/cm}$. 
Here, we must emphasize that this sub-natural linewidth can only be achieved due to the collective superradiant nature of the system, and it occurs even in the presence of the inhomogeneous broadening on the Hz level shown in Fig.~\ref{fig:10}(g). 
This suppression of noise is on the order of one million and can be only explored in a truly continuous system without any Fourier limitations.

If we consider the sensitivity of the output frequency to variations of the atomic flux $\Phi$ by looking at Fig.~\ref{fig:10}(b), we see that this sensitivity is minimized for negative values of $G_B$ between $-120\,{\rm mG/cm}$ and $-90\,{\rm mG/cm}$.
In this region, the shift sensitivity is on the order of $2\pi \times 0.3~{\rm Hz}$ per $10^6$~atoms/s.
Thus, a mean atomic flux of $\Phi=7\times 10^6~{\rm s^{-1}}$ with 5\% atom number fluctuation would lead to about a 100~mHz broadened linewidth of the output frequency. These fluctuations seem to be the main factor limiting the short-term frequency stability of the output laser signal.

\begin{figure*}
    \centering
    \includegraphics[width=1\textwidth]{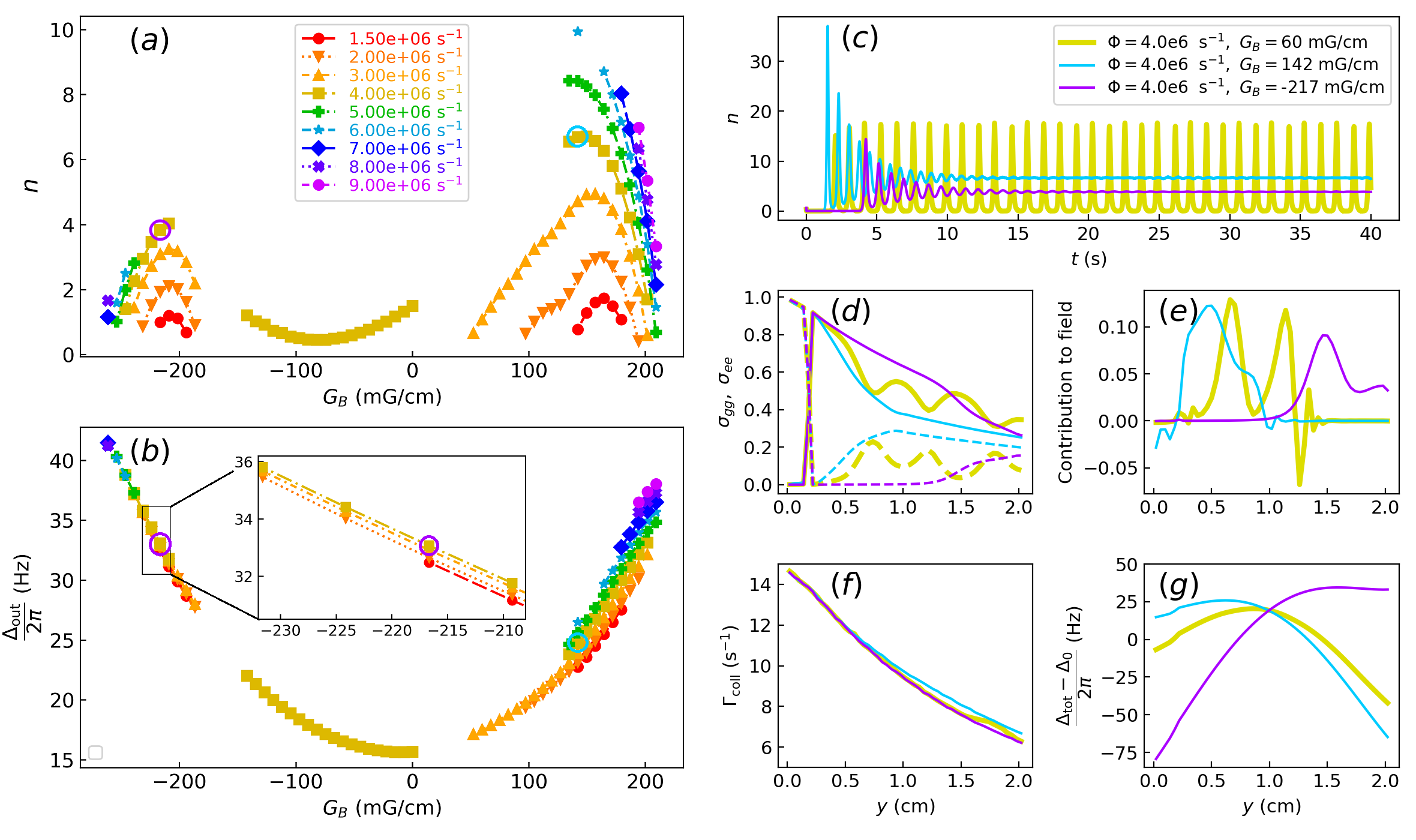}
    \caption{Simulation for the extended dephasing model. (a) Intracavity photon numbers and (b) frequency shifts of the output field (relative to the atomic transition) are shown for different magnetic field gradients $G_B$ and atomic fluxes $\Phi$ (see legend; same color code for (a) and (b)). Inset: magnetic field gradient range in which output frequency is least sensitive to flux fluctuations. Circled points correspond to stable solutions investigated in detail in (c) -- (g). (c) Examples of intracavity photon number over time for $\Phi=4\times 10^6~{\rm s^{-1}}$ and three different values of the magnetic field gradient $G_B$. (d) Position dependent distributions of populations $\sigma_{ee}$ (solid) and $\sigma_{gg}$ (dashed)  along the conveyor. (e) Atomic position dependent contribution to the cavity field $a$ defined as the imaginary part of $\sigma_{ge} \exp[-i{\rm arg}(a)]$. (f) Distribution of collision-induced dephasing. (g) Total (magnetic plus collision-induced) shift $\Delta_{\rm tot}=\Delta(y)+\Delta_{\rm coll}$ along the conveyor. The color code for (c) -- (g) is shown in (c). The magnetic field is set to $B_0=574~{\rm G}$ and all other parameters are the same as in Figure~(\ref{fig:10})}
    \label{fig:11}
\end{figure*}
\subsection{Simulation with extended dephasing model}
\label{sec:las:full}
Our simulations thus far are based on the model presented in Ref.~\cite{Lisdat09}, where the authors assume that the main source of dephasing is elastic collisions with atoms in the ${\rm ^1S_0}$ ground state. According to that model, the ground state dephasing coefficient due to elastic collisions, $\gamma_{\rm dep}$, is much larger than the loss coefficients due to inelastic collisions between the atoms in the excited state ($\gamma_{ee}$) and inelastic collisions between ground and excited state atoms ($\gamma_{ge}$), by a factor of about 100 and 1000, respectively (see Section~\ref{sec:las:model}). The role of dephasing in elastic collisions with atoms in the excited state has thus far been ignored.

However, we start with a fully inverted sample of atoms, which means that at the beginning of the emission zone, most atoms are in the excited state. 
Furthermore, the atoms will spend a considerable time in the excited state within the emission zone, which can be seen in Figure~\ref{fig:10}(d). 
Therefore, dephasing due to elastic collisions in the excited state cannot be ignored in our case. 
In this section, we investigate the feasibility of the bad cavity laser with an ``extended'' dephasing model, where we now include dephasing due to elastic collisions in the excited state. 
Due to the lack of experimental data, we set the dephasing coefficient for the excited state $\gamma_{\rm dep}=3.2\times10^{-10}~{\rm cm^3/s}$ equal to the dephasing coefficient in the ground state. Its accurate determination is left as the subject of future experimental study. 
It must be noted that dephasing due to elastic collisions between excited and ground state atoms is still not considered in this model.

To now account for the additional two-body collision dephasing effects, we use the following expression for the dephasing rate:
\begin{align}
\Gamma^j_{\rm coll}&=n_j \left[\frac{\sigma^j_{ee} \gamma_{ee}+ [\sigma^j_{gg}+\sigma^j_{ee}] \gamma_{ge}}{2} + \gamdep (\sigma^j_{gg}+\sigma^j_{ee})\right].
\label{eq:conv:23}
\end{align}
With this new dephasing rate, we find that we must increase the bias magnetic field in order to achieve a larger atom-field coupling to obtain steady-state superradiant emission. 
Here, we choose $B_0=574~{\rm G}$, which is 2.5 times stronger than before. Calculations with COMSOL show a nearly proportional scaling of magnetic field deviations $B(y)-B_0$, within 1\%. This field corresponds to $\gamma_{\rm {^3P_0}\rightarrow{^1S_0}}=6.43 \times 10^{-4}~{\rm s^{-1}}=2\pi \times 102~{\rm \mu Hz}$. 

The dependences of the intracavity photon number and the shift of the output radiation on $G_B$ for stable solutions are presented in Figure~\ref{fig:11}(a,b). We see that in contrast to the dependence presented in Figure~\ref{fig:10}(a), the intracavity photon number as a function of $G_B$ has two maxima and a wide dip between $-180<G_B<50\,\rm mG/cm$, caused by the proportionally stronger variation of the magnetic shift along the optical conveyor.  

In Figure~{\ref{fig:11}}(c-g), we present results of simulations for $\Phi=4\times 10^6~{\rm s^{-1}}$ and three different values of the magnetic field gradient, similar to how it has been done for Figure~\ref{fig:10}.
We use a smaller atomic flux than was used in the basic dephasing model since no stable solutions are obtained for higher values of flux with the extended model.
Again, we present two stable (violet and light blue curves) solutions and one unstable (yellow curve) solution. The variation of the position-dependent shift presented in Figure~\ref{fig:11}(g) is much larger than in Fig.~\ref{fig:10}(d) because of the proportionally larger inhomogeneity of the magnetic field along the optical conveyor. The maximum steady-state intracavity photon number is $n\approx 6.62$ in Fig.~\ref{fig:11}(c), which is 45\% lower than the one presented in Fig.~\ref{fig:10}(c). This can be partially explained by the 40\% lower atomic flux. For such a flux, the maximum power that can be emitted into the cavity mode is $\hbar \omega \Phi \approx 1.14\,{\rm pW}$, whereas the power transferred into the cavity field is about $P_{\rm field}= 350$\,fW for $G_B=142\,{\rm mG/cm}$, corresponding to a 31\% transfer efficiency. As a result, the output laser power $P_{\rm out}=\eta P_{\rm field}$ can be estimated as about 90\,fW, again by assuming all mirrors of the cavity have the same reflectivity. 

In contrast to the situation presented in Figure~\ref{fig:10}~(c), the unstable solution produces relatively regular pulses, rather than a chaotic regime. 
The dynamics of intracavity populations presented in Figure~\ref{fig:11}(d) demonstrate nearly the same population transfer efficiency for $\Phi=4.0\times 10^6\,{\rm s^{-1}}$, $G_B=142~{\rm mG/cm}$ (light blue curves) as for $\Phi=7.0\times 10^6\,{\rm s^{-1}}$, $G_B=0$ in the basic model (black curves in Figure~\ref{fig:10}(d)). The smaller atomic number density and relatively smaller collision losses used in the extended dephasing model lead to the same transfer efficiency, even though the total shift in the emission zone is larger. 
The atomic contribution into the intracavity field for the unstable solution (yellow curve in Figure~\ref{fig:11}(e)) has 2 strong peaks corresponding to simultaneous lasing on two slightly different frequencies, resulting in pulses. 
The collision-induced dephasing rate presented in Figure~{\ref{fig:11}}(f) is nearly proportional to the total population change, whereas in our basic model, it is primarily determined by the population of the ground state.

The simulations presented in Figure~\ref{fig:11} were performed for a pumping-induced light shift $\delta_p$ of zero. To check the robustness of the output laser signal against a non-zero shift, we performed simulations for $\delta_{p}^0=2\pi \times 25 \,{\rm Hz}$ and $\delta_{p}^0=2\pi \times 500\,{\rm Hz}$. 
Using these two values for the pumping-induced light shift to simulate the generated field for the two marked solutions in Figure~{\ref{fig:11}}(c), we achieve a stable solution in all four cases. 
For those cases, the variation of the number $n$ of intracavity photons is less than 1\%, and the constant output frequency shift $\Delta_{\rm out}$ was also smaller than $2\pi \times 50~{\rm mHz}$. This leads us to the assumption that the pumping zone has a very small influence on the performance of the superradiant laser. 
 
We also estimate the minimum achievable linewidth for such a system for two selected points corresponding to a stable solution. For $\Phi=4\times 10^6~{\rm s^{-1}}$ and $G_B=142\,{\rm mG/cm}$ the estimated linewidth of the output radiation is on the level of $2 \pi \times 120~{\rm \mu Hz}$, and for $\Phi=4\times 10^6~{\rm s^{-1}}$ and $G_B=-217\,{\rm mG/cm}$, it is on the level of $2\pi \times 50~{\rm \mu Hz}$. The larger values of the linewidth in comparison to the ones presented in the previous section are due to the larger variations of the position-dependent shift along the optical conveyor. In this case, the collective nature of superradince again leads to a suppression of the inhomogeneous broadening effects in our system on the order of one million, resulting in a linewidth comparable to the natural linewidth at this large bias magnetic field.

For $G_B$ lying between approximately $-230~{\rm mG/cm}$ and $-210~{\rm mG/cm}$ Figure~{\ref{fig:11}} (b), we see that the output frequency is more robust against variations in the atomic flux $\Phi$ than in the case considered in Section~\ref{sec:las:results}. Five percent fluctuations of the atom number around $\Phi=4\times 10^6\,{\rm s^{-1}}$ lead to a broadening of about 50\,mHz, which corresponds roughly to broadening by a factor of one thousand compared to the minimum achievable linewidth. We can conclude that in our simulations, atom number fluctuations remain the main source of instability for the superradiant laser.
\section{Outlook}

In this paper, we have focused on the ${\rm ^{88}Sr}$ isotope due to its high natural abundance and its simple internal structure.  However, it requires a strong external magnetic field to partially allow the \transclock transition, which directly leads to unavoidable position-dependent shifts due to imperfections in the applied field. Additionally, the strong s-wave collisions between bosonic atoms cause significant dephasing and shifts. As we have seen, atom number fluctuations are the main linewidth broadening mechanism in our system. Therefore, reducing these fluctuations would be the most straightforward path to improving the frequency stability of the superradiant laser. More accurate experimental measurements of dephasing and loss coefficients in the ground and excited state, as well as collisional-induced shifts could also improve our understanding of the system and allow us a more quantitative numerical optimization of parameters to minimise linewidth broadening effects. 

Our simulations also indicate that light shifts associated with pumping have a very weak influence on the output frequency because the atoms do not contribute to the cavity field while being pumped due to strong dephasing. 
Therefore, we could include one more repumping zone in the optical conveyor, with the purpose of repopulating the excited state, to potentially increase the emitted power.

Alternatively, we could explore fermionic $^{87}$Sr on the $^1$S$_0,F=9/2, m_F= \pm9/2 \rightarrow{{\rm ^3P_0},F=9/2, m_F=\pm9/2}$ transition in our system. The more complex internal structure, which includes hyperfine and Zeeman splitting, would lead to more complicated cooling and pumping schemes compared with the ones presented above. However, the non-zero clock transition rate and the resulting stronger coupling to the cavity, even at zero magnetic field, leads to a lower collective atomic number threshold for superradiant emission. Together with the suppression of $s$-wave collisions due to the Pauli exclusion principle, one could expect better robustness of the output laser field against fluctuations of the magnetic field or the atomic flux. This investigation is left to future theoretical and experimental study.
\section{Conclusion}

In this paper, we consider the design of a continuously operating superradiant laser on the \transclock transition in ${\rm ^{88}Sr}$.
We discuss the mechanism by which we will continuously load atoms from the dipole trap into a magic wavelength optical conveyor lattice generated inside a bow-tie cavity. This bow-tie cavity creates the strong collective coupling between atoms that should enable superradiant emission. 
We have also simulated highly efficient atom loading for a moving optical lattice with a speed of a few cm per second. We showed that up to 83\% of atoms in the dipole trap get trapped and pumped in the optical lattice with an average energy slightly above $16~{\rm \mu K}$.

Next, we numerically simulated the output of a continuous superradiant laser, taking into account the inhomogeneity of the magnetic field, collisional dephasing, shifts and losses, as well as pumping-induced effects. For collisional decoherence, we considered two models. 
In the first model, adapted from \cite{Lisdat09}, we supposed that ground state collisions are the main source of decoherence. We show that with experimentally realistic parameters, we can achieve superradiant lasing with an output power of about $160~{\rm fW}$ and a quantum fluctuation-limited linewidth on the order of a few ${\rm \mu Hz}$. 
The main limitation on the linewidth appears to be broadening due to fluctuations of the atomic flux. This broadening is determined by the collisional shift, as well as by the redistribution of the atomic coherence over the emission zones.
For the used flux, a five percent atom number fluctuation would broaden the linewidth to about $100\,{\rm mHz}$.
In the extended dephasing model, we included the contribution of the excited state atom collisions to dephasing. We showed that stable superradiant lasing becomes possible, but with a higher magnetic field of about 600\,G, which leads to stronger variations of the magnetic field and position-dependent shifts in the emission zone. 
We have added an extra magnetic field gradient to shape the magnetic field to partially compensate these variations. The larger inhomogeneous broadening effects lead to a quantum fluctuation-limited linewidth of about $100\,{\rm \mu Hz}$ and an output power of around $5-10~{\rm fW}$. As with the basic model, the main broadening mechanism is fluctuations of the atomic flux. Five percent fluctuation of the flux would lead to about a $50\,{\rm mHz}$ broadened linewidth in the experiment. For both models, we have shown that the effect of the light shift caused by the pumping fields is nearly negligible for the generation of superradiant emission because pumped atoms experience strong dephasing, which destroys any correlations between the ground and excited states, and the light shift protects the cavity field from the interaction with these atoms.

From this study, we can conclude that continuous superradince on the sub-mHz transition is possible and that such a system should be competitive with today's state-of-the-art short-term references, even taking into account the main broadening effects.

\begin{acknowledgments}
We thank Francesca Fam\`{a} and Camila Beli Silva for their contributions to the experimental apparatus and Ronald Hassing for his support with the magnetic field simulations. This work has received funding from the European Union’s (EU) Horizon 2020 research and innovation program under Grant Agreement No. 820404 (iqClock project), No. 860579 (MoSaiQC project) and No. 856415 (Thorium nuclear clock). It further received funding from the Dutch National Growth Fund (NGF), as part of the Quantum Delta NL programme. The simulation scripts have been written in Julia programming language \cite{Julia}, partially using QuantumOptics.jl package \cite{JuliaQuantumOptics}. The computational results presented have been achieved in part using the Vienna Scientific Cluster (VSC). 

\textit{Author contributions---} This work was proposed by GK, SD, SAS, SB, and FS. The theoretical analysis was conducted by SD and GK, with experimental guidance from BH, AS, SB, and SZ. The analysis was based on the experimental apparatus designed by SB, with significant contributions from SZ and FS. The manuscript was drafted by GK and SD, and was substantially edited by BH and AS. All authors reviewed the manuscript. Theoretical work was supervised by GK. GK and FS secured the project funding.

\end{acknowledgments}


\appendix
\section{Polarizibilities and Design Parameters}
\label{App:a}

We have calculated the polarizibility using the following expression
\begin{equation}
\begin{split}
\alpha_i^{E_1}=&\alpha_i^{(0)}+i(\epsilon^*\times\epsilon)_z\cdot \frac{m}{F}\cdot\alpha_i^{(1)}\\
&+\frac{(3\epsilon^*_z\epsilon_z-1)(3m^2-F(F+1))}{2F(2F-1)}\alpha_i^{(2)}.
\end{split}
\label{app:P}
\end{equation}
The three components are scalar, vector, and tensor polarizabilities:
\begin{equation}
\begin{split}
\alpha_i^{(0)}=&\beta_{0}\sum_{\mathbf{\Vec{n}'F'}}|\langle{\mathbf{\Vec{n}}F}||\hat{d}||{\mathbf{\Vec{n}}'F'}\rangle|^2
\frac{2\omega_{\mathbf{\Vec{n}}'F',\mathbf{\Vec{n}}F}}{\omega_{\mathbf{\Vec{n}}'F',\mathbf{\Vec{n}}F}^2-\omega^2},
\\
\alpha_i^{(1)}=&\beta_{1}\sum_{\mathbf{\Vec{n}'F'}}|\langle{\mathbf{\Vec{n}}F}||\hat{d}||{\mathbf{\Vec{n}}'F'}\rangle|^2
\\
&\times \frac{\omega}{\omega_{\mathbf{\Vec{n}}'F',\mathbf{\Vec{n}}F}^2-\omega^2}(-1)^{F+F'}
\left\{\begin{array}{ccc}
  1 & 1 & 1 \\
  F & F & F'
\end{array}\right\},\\
\alpha_i^{(2)}=&\beta_{2}\sum_{\mathbf{\Vec{n}'F'}}|\langle{\mathbf{\Vec{n}}F}||\hat{d}||{\mathbf{\Vec{n}}'F'}\rangle|^2 \\
&\times \frac{\omega_{\mathbf{\Vec{n}}'F',\mathbf{\Vec{n}}F}}{\omega_{\mathbf{\Vec{n}}'F',\mathbf{\Vec{n}}F}^2-\omega^2}(-1)^{F+F'}
\left\{\begin{array}{ccc}
  1 & 1 & 2 \\
  F & F & F'
\end{array}\right\},\\
\end{split}
\end{equation}
with 
\begin{equation}
\begin{split}
&\beta_{0}=\frac{1}{3\hbar(2F+1)}, \quad \beta_{1}=\frac{\sqrt{6F}}{\hbar\sqrt{(2F+1)(F+1)}},\\
&\beta_{2}=\frac{1}{\hbar}\sqrt{\frac{40F(2F-1)}{(2F+3)(2F+1)(F+1)}}.
\end{split}
\end{equation}
Here the sums are taken over the atomic states (denoted by the prime index) which have $E1$ coupling to the state $i$, $\mathbf{\Vec{n}}=\{n,S,J,L\}$ represent the quantum numbers of the atomic level: $n$ is the atomic principal quantum number, $S$ is the electronic spin, $L$ is the electronic orbital angular momentum, $J$ is the total angular momentum of electronic shells; $F$ is the total momentum of the atom, and $\omega_{\mathbf{\Vec{n}}'F',\mathbf{\Vec{n}}F}=\omega_{\mathbf{\Vec{n}}'F'}-\omega_{\mathbf{\Vec{n}}F}$ is the transition frequency. Also, we use here the following convention for the square of reduced matrix element:
\begin{equation}
|\langle{\mathbf{\Vec{n}}F}||\hat{d}||{\mathbf{\Vec{n}}'F'}\rangle|^2
=\frac{3 \hbar c^3 \gamma_{\mathbf{\Vec{n}'}F',\mathbf{\Vec{n}F}} (2F^{\rm up}+1)}{4|\omega_{\mathbf{\Vec{n}'}F',\mathbf{\Vec{n}F}}|^3},
 \label{eq:A1:redME}
\end{equation}
where $\gamma_{\mathbf{\Vec{n}'}F',\mathbf{\Vec{n}F}}$ is the spontaneous transition rate between the respective hyperfine sublevels (${\rm ^{88}Sr}$ does not have a hyperfine structure, therefore, $F=J$, and $\gamma_{\mathbf{\Vec{n}'}F',\mathbf{\Vec{n}}F}=\gamma_{\mathbf{\Vec{n}'},\mathbf{\Vec{n}}}$.
\begin{equation}
F^{\rm up}=\left\{ 
\begin{array}{cl}
F', & \omega_{\mathbf{\Vec{n}'}F',\mathbf{\Vec{n}}F}>0, \\
F, & \omega_{\mathbf{\Vec{n}'}F',\mathbf{\Vec{n}}F}<0.
\end{array}
\right.
\end{equation}
For calculation of the polarizability we used the data presented in \cite{Zhou10}.

In Table~\ref{table:DP} we present potentials and geometrical parameters of potential beams, namely the dipole guide, the reservoir beam, and the optical conveyor.
\begin{table}
\caption{Parameters for potential beams.}
\centering
\begin{tabular}{p{2.7cm}>{\centering\arraybackslash}p{1.5cm}>{\centering\arraybackslash}p{1.5cm}>{\centering\arraybackslash}p{1.5cm}} 
\hline\hline
Beam Type & Dipole guide & Reservoir & Conveyor \\ \hline
Propagation along & $z$-axis &  in $x$-$z$ plane  $5^{\circ}$~from z& $y$-axis \\
Polarization & along $x$  & $\sim 85^\circ$ from $x$ & $\sim 74^\circ$ from $x$ \\ 
$U_0({\rm ^1S_0}),~{\rm \mu k}$  & $166$ & $50$ & $30$ \\
$U_0({\rm ^3P_{1},m=0}),~{\rm \mu k}$  & $105.4$ & $31.97$ & $37.65$ \\
$U_0({\rm ^3P_{1},m=1}),~{\rm \mu k}$  & $129.62$ & $38.93$ & $30$ \\
Waist, $\rm \mu m$ & $(200,200)$ & $(400,100)$ & $140 $ \\
Wavelength, $\rm nm$ & $1070$ & $1070$ & $813$ \\
\hline\hline
\end{tabular}
\label{table:DP}
\end{table}

\section{Semiclassical Monte Carlo simulation method for atomic cooling and pumping}
\label{App:B}

\begin{table}
\centering
\caption{Parameters of molasses beams. Here, $\Gamma$ is the spontaneous rate of ${\rm {^3P_1} \rightarrow {^1S_0}}$ transition, $\Delta^{\rm D(max)}_{\rm 2}=2\pi\times1.262~{\rm MHz}$, $\Delta^{\rm D(max)}_{\rm 3}=2\pi\times0.757~{\rm MHz}$ and $\Delta^{\rm R(max)}_{\rm 3}=2\pi\times0.230~{\rm MHz}$ are the maximal light shifts associated with dipole guide and reservoir beams respectively (see expression~(\ref{eq:appB:Delta_DR})), and $\Delta^{\rm D+R~(max)}_{\rm 3}=\Delta^{\rm D(max)}_{\rm 3}+\Delta^{\rm R(max)}_{\rm 3}$.}
\begin{tabular}{p{1.7cm}>{\centering\arraybackslash}p{1.98cm}>{\centering\arraybackslash}p{1.98cm}>{\centering\arraybackslash}p{1.98cm}} 
\hline\hline
Molasses Beam & M1 & M2 & M3 \\ \hline
Intensity (total) & $ 2\, I_{\rm sat}$ & $I_{\rm sat}$ & $ I_{\rm sat}$ \\ 
$\Delta_{\rm band}/(2\pi)$ & $\rm 10~kHz$ & $\rm 10~kHz$ & $\rm 10~kHz$ \\ 
$N_f$ & 12 & 4 & 5 \\ 
$\rm \delta_i^c$ & $-45\frac{\Gamma}{2}+\Delta^{\rm D(max)}_{\rm 2}$ & $-15\frac{\Gamma}{2}+\Delta^{\rm D+R~(max)}_{\rm 3}$ & $-15\frac{\Gamma}{2}+\Delta^{\rm D+R~(max)}_{\rm 3}$ \\ 
Waist, $\rm \mu m$ & (200,200) & (200,200)  & (200,200) \\ 
$\rm \lambda,~\rm nm$ & 689 & 689 & 689 \\ \hline
center, $~\rm (mm)$ & (0,0,  -5) & $(0,0,0)$ & $(0,0,0)$ \\ 
Polarization & in $y-z$ plane & in $y-z$ plane & along $y$ \\ 
Propagation  & in $y-z$ plane, $7^{\circ}$ from $z$ & in $y-z$ plane $7^{\circ}$ from $z$ & along $x$\\ \hline\hline
\end{tabular}
\label{table:PMB}
\end{table}

To simulate deceleration and cooling of the atoms by the molasses beams we employ a ``semi-classical Monte Carlo'' model, where we suppose that an atom will not significantly change its position during the typical internal state evolution time.
Therefore, the internal state of an atom depends only on its instantaneous position and velocity, not on the previous history.
The position-dependent scattering rate associated with the molasses beams is given by the following equation:
\begin{align}
\Gamma_{i}(r,v)=\sum_{k}\bigg(\frac{\gamma_{3g} s_{ik}(r,v)}{2+s_{ik}(r,v)}\bigg)\bigg(1+\sum_{i,j}\frac{s_{ij}(r,v)}{2+s_{ij}(r,v)}\bigg)^{-1},
\label{app:sr}
\end{align}
where $i$ represents the substates $^3P_1,m=\{-1,0,1\}$ corrsponding to $i=\{1,2,3\}$ respectively, $\gamma_{3g}$ is the spontaneous decay rate of the ${\rm {^3P_1}\rightarrow {^1S_0}}$ transition, and $k$ is associated with different frequencies in the band. We are specifically utilizing $i=\{1,2\}$ transitions. The associated saturation parameter and detuning are given by
\begin{align}
\begin{split}
&s_{ik}(r,v)=\frac{I^i_k(r)}{4I_{\rm sat}^i}\bigg(1+\bigg(\frac{2\delta_{ik}(r,v)}{\gamma_{3g}}\bigg)^2\bigg)^{-1}\\
&\delta_{ik}(r,v)=\delta_{ik}'-\Delta^{\rm D}_{i}(r)-\Delta^{\rm R}_{i}(r)+\Vec{k}_{ik}\cdot\Vec{v},\\
&I^i_{\rm sat}=\frac{2\pi^2\hbar c \gamma_{3g}}{3\lambda^3}\quad, \quad \delta_{ik}'=\frac{\Delta_{\rm band}(2k-N_f-1)}{2(N_f-1)}+\delta_i^c\\
\end{split}
\label{app:sd}
\end{align}
$\Delta_{\rm band}$ is the bandwidth, $\delta_i^c$ is the the central frequency and $N_f$ is the total number of frequency. For $i$th transition, the differential light shifts associated with dipole guide beam ($\Delta^{\rm D}$), reservoir ($\Delta^{\rm R}$), and optical lattice ($\Delta^{\rm OL}$) are given by 
\begin{equation}
\Delta^{\rm \{D,R\}}_i(r)=-\frac{1}{\hbar}\bigg(\frac{\alpha_{\rm ^1S_0}-\alpha_{i}}{\alpha_{\rm ^1S_0}}\bigg)U_{\rm \{D,R\}}^{{}^1S_0}(r),
\label{eq:appB:Delta_DR}
\end{equation}
\begin{equation}
\Delta^{\rm OL}_i(r,t)=-\frac{1}{\hbar}\bigg(\frac{\alpha_{\rm ^1S_0}-\alpha_{i}}{\alpha_{\rm ^1S_0}}\bigg)U_{\rm OL}^{\rm ^1S_0}(r,t).
\label{app:ss_ol}
\end{equation}

\begin{table}
\caption{Parameters of pumping beams. Here P-1 acts on the $\ket{\rm ^1S_0} \rightarrow \ket{\rm ^3P_1, m=-1}$  transition. Similarly, P-3 and P-4 acts on the $\ket{\rm ^1S_0} \rightarrow \ket{\rm ^3P_1, m=\mp 1}$ transitions, whereas P-5, P-6, and P-7 pump atoms from the $\ket{{\rm ^3P_2},m=-1}$, $\ket{{\rm ^3P_2} ,m=0}$ and $\ket{{\rm 
^3P_2},m=1}$-states to the $\ket{{\rm ^3S_1} ,m=0}$- state, respectively. The partial decay rates $\Gamma$ associated with the corresponding transitions are $\Gamma_{\rm total}$ for P-1, $0.5\, \Gamma_{\rm total}$ for P-3,3, $0.3\, \Gamma_{\rm total}$ for P-5,7, and $0.4\, \Gamma_{\rm total}$ for P-6 respectively. The Intensities mentioned in the table corresponds to the circular polarized field component.}
\centering
\begin{tabular}{p{1.9cm}>{\centering\arraybackslash}p{1.9cm}>{\centering\arraybackslash}p{1.9cm}>{\centering\arraybackslash}p{2cm}} 
\hline\hline
Pumping Beam & P-1 & P-\{3,4\} & P-\{5,6,7\} \\ \hline
Intensity & $ 0.32\, I_{\rm sat}$ & $\{0.5,0.02\}\, I_{\rm sat}$ & $ 0.02\, I_{\rm sat}$ \\
$\rm \Gamma_{total},~\mu s^{-1}$ & $4.69\times 10^{-2}$ & 27.0& $\rm 42.0$ \\
Waist, $\rm \mu m$ & (250,250) & (250,250)  & (250,250) \\
$\rm \lambda,~\rm nm$ & 689 & 688 & 707 \\
center, $~\rm (mm)$ & (0,2,0) & (0,2,0) & (0,2,0) \\
Polarization & along x-axis & along x-axis & along \{x,z,x\}-axis\\ \hline
Propagation  & along z-axis & along z-axis & along z-axes\\
$\Delta/(2\pi)$  & $-1$ kHz & $\{0,-2\}$ MHz & $\{-19.9,0.1,$ $20.1\}$ MHz\\ \hline\hline
\end{tabular}
\label{table:PMP}
\end{table}

We use the position-dependent scattering rate to generate random numbers and  artificially model the absorption and emission of photons by evolving the scattering rate (\ref{app:sr}) along the trajectory of atoms. We perform the simulation by following these steps:
\begin{itemize}
  \item Starting with an initial state $(\vec{x}_i,\vec{p}_i)$ at time $t_i$ we evolve the equation of motion $(\dot{\vec{x}}_i,\dot{\vec{p}}_i)$ and the scattering rate $\Gamma_m(x,v)$ till $t_{i+1}=t_i+dt$.
  \item Generate a vector $\Vec{v}_s=(\sin{\theta}\cos{\phi},\sin{\theta}\sin{\phi},\cos{\theta})$ with $\theta$ and $\phi$ randomly generated numbers and a random number $r_1$ in the interval $[0,\pi]$, $[0,2\pi]$ and $[0,1]$ respectively.
  \item If $r_1>e^{-\int_{t_i}^{t_{i+1}}\Gamma_m(\vec{x}_i,\vec{v}_i)dt}$ set $\Vec{p}_{i+1}= \Vec{p}_i-\hbar \Vec{k} +\Vec{v}_s\hbar k$ and regenerate $\{r_1, \theta, \phi\}$ or else without changing anything take another step.
  \item Repeat the above three steps until we reach the desired evolution time.
\end{itemize}
In our system, we utilize a broad spectrum of frequencies to address a variety of velocities. The maximum scattering rate is significantly lower than the spontaneous emission rate, allowing us to assume that an atom quickly emits a photon spontaneously after absorbing it.

For the simulation of pumping, we employed the master equation detailed in Appendix C. We identified a set of detunings where the trapping of population into the dark states doesn't play a significant role. Leveraging this carefully selected set of parameters, as described in Table~\ref{table:PMP}, we found that the population probability of atoms in states $^3P_1$, $^3P_2$, and ${\rm ^3S_1}$ remain low. The light field intensity for the transition $\ket{3} \rightarrow \ket{8}$ is chosen such that $\Gamma^{P}_{38} \gg \Gamma^{\text{\rm decay}}_{31}$, thereby preventing excessive scattering events from the $\ket{1} \leftrightarrow \ket{3}$ transition, which would otherwise cause heating.

Under conditions of far-detuned lasers, the intensities given in Table~\ref{table:PMP}, and the fact that the motion of these atoms does not significantly change during the pumping cycle, we can assume that only the effect of individual lasers acting on two levels, at given point in time, needs to be considered. To estimate the pumping efficiency and the heating of atoms within the moving optical lattice, we apply the SCMC method described earlier. The pumping rate associated with each individual two-level system is calculated using:
\begin{align}
\Gamma^{p}_{m}(\vec{r},\vec{v})=\frac{\displaystyle{\sum_{j}\frac{\gamma_{3g} s^p_{mj}(\vec{r},\vec{v})}{2+s^p_{mj}(\vec{r},\vec{v})}}}
{1+\displaystyle{\sum_j\frac{s^p_{mj}(\vec{r},\vec{v})}{2+s^p_{mj}(\vec{r},\vec{v})}}},
\label{app:sr}
\end{align}
 where 
\begin{align}
s^p_{mj}(\vec{r},\vec{v})=\frac{I^m_j(r)}{4I_{\rm sat}^m}\bigg(1+\bigg(\frac{2\delta_{mj}(\vec{r},\vec{v})}{\gamma_{3g}}\bigg)^2\bigg)^{-1}.
\end{align}
The spontaneous decay from ${\rm ^3S_1}$ is modeled by generating a random number based on the relative decay rates for different transitions. Momentum kicks are applied similarly to the molasses simulation. The decay rate from ${\rm ^3S_1}$ is 100 times higher than any other process. For the ${\rm ^3P_1}$ state, which has a smaller decay rate, the momentum kicks are modeled using Einstein's rate equation. Random numbers $r_1$, $r_{2}$, and $r_3$ are generated in the interval $[0,1]$ and compared with the probabilities associated with different channels.

If the system is in state $\ket{1}$ and $r_1 > e^{-\int_{t_i}^{t_{i+1}}\Gamma_{13}^p(x,v)dt}$, the state is set to $\ket{3}$. Then, depending on whether $r_2 > e^{-\int_{t_i}^{t_{i+1}}\Gamma_{38}^p(x,v)dt}$ or $r_3 > e^{-\int_{t_i}^{t_{i+1}}\Gamma_{31}^{\rm decay}dt}$, the state is either changed to $\ket{8}$ or returned to $\ket{1}$, with corresponding momentum kicks applied.

\section{Reduction of multilevel repumping scheme to an effective 2-level scheme}
\label{App:C}

\begin{figure}
    \centering
    \includegraphics[width=0.45\textwidth]{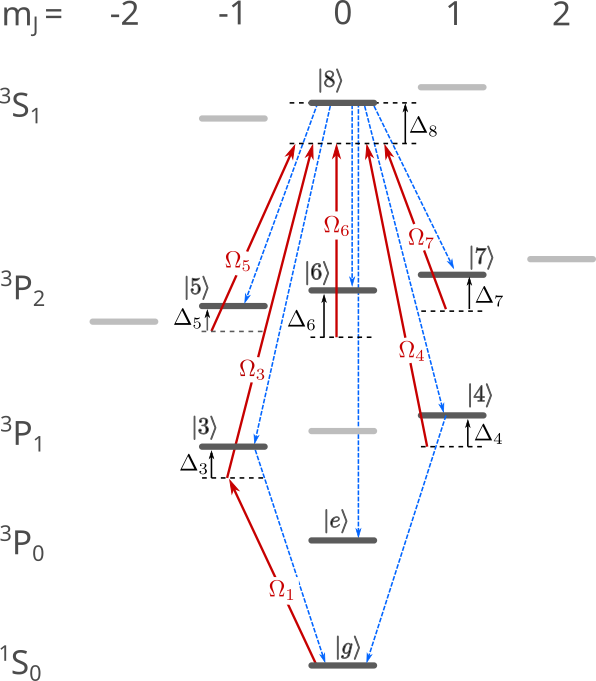}
    \caption{Pumping scheme with relevant matrix elements $\Omega_j$ and frequency detunings $\Delta_j$. Laser-induced transitions are shown by solid red, and spontaneous decays by dashed blue arrows. The levels which do not participate in the repumping process are shown in pale gray.  }
    \label{fig:C:PumpParameters}
\end{figure}

In this section we specify parameters of the realistic multilevel pumping scheme and perform a mapping of this pump scheme to an effective 2-level incoherent pumping scheme, following the method developed in \cite{Hotter21repump}. Here we neglect all the collision-induced processes, because the pumping occurs in a very narrow zone (250~nm waist), and these process will not  significantly affect the internal state of the atoms. Therefore, we consider isolated atoms interacting with pumping fields. The multilevel pumping scheme is presented in Figure~\ref{fig:C:PumpParameters}. Here, we specify the notation for sublevels, matrix elements and frequency detunings. 

The Hamiltonian for such system in resonant approximation can be written as
\begin{align}
\frac{\HH^0}{\hbar}&=\sum_{j=g}^8\omega_j\sig_{jj} 
    + \Omega_{1}\left[\sig_{g3} e^{i(\omega_{3g}^Lt+\phi_{1})}
                      +\sig_{3g} e^{-i(\omega_{3g}^Lt+\phi_{1})}\right] \nonumber \\
    &+\sum_{j=3}^7 \Omega_{j}\left[
                    \sig_{j8} e^{i(\omega_{8j}^Lt+\phi_{j})}
                    +\sig_{8j} e^{-i(\omega_{8j}^Lt+\phi_{j})}
                  \right].
\label{eq:C:H0}
\end{align}
Here $\sig_{ij}=\ket{i}\bra{j}$, $\Omega_j$ are transition matrix elements which can be expressed via intensities $I_j$ of the respective laser fields, saturation intensities $I^j_{\rm sat}$ and the respective spontaneous transition rates $\gamma_{kj}$ as
\begin{equation}
    \Omega_j=\sqrt{\frac{I_j}{8I_{\rm sat}^j}}\gamma_{kj}.
    \label{eq:C:OmegaVsIsat}
\end{equation}
Therefore, $\omega_{ij}^L$ is the frequency of the laser acting on the $\ket{j}\rightarrow \ket{i}$ transition, and $\phi_{i}$ is the time-dependent phase of a laser applied to $\ket{g}\rightarrow\ket{3}$ transition at $i=1$, as well as $\ket{i}\rightarrow\ket{8}$ transition at $i=3$ to 7. Here we suppose that all the lasers are independent, and their fluctuations corresponds to a white frequency noise, namely
\begin{equation}
\langle \dot{\phi}_{i}(t)\dot{\phi}_{j}(t')\rangle=\Gamma^L_{i}\delta_{ij} \delta(t-t'),
\label{eq:C:fluct}
\end{equation}
where $\Gamma^L_{i}$ is the linewidth of the laser acting on the $\ket{i}\rightarrow\ket{j}$ transition. 

\begin{table*}
\caption{Dissipative processes in the 8-level pumping scheme. Decay rates were calculated as $\gamma_{\ket{n'L'J'm'}\rightarrow\ket{nLJm}}=\gamma_{\ket{n'L'J'}\rightarrow\ket{nLJ}} (C^{J'm'}_{Jm1m'-m})^2$, where the decay rates $\gamma_{\ket{n'L'J'}\rightarrow\ket{nLJ}}$ between fine-structure levels $\ket{n'L'J'}$ and $\ket{nLJ}$ were taken from \cite{NIST}, $C^{J'm'}_{Jm1m'-m}$ are Clebsch-Gordan coefficients, and $J',m'\,(J,m)$ are the angular momentum and its projection associated with the upper (lower) state. The value of $\gamma_{eg}=\gamma$ depends on the applied bias magnetic field and does not affect other parameters of the equivalent 2-level scheme.}
\centering
\begin{tabular}{ccccc}
\hline\hline
$j\vphantom{\int}$& $\hat{J}_j$   & $R_j$ & value (order of magnitude) & description \\ \hline
1  & $\sig_{58}$   & $\gamma_{85}$ & $1.26\times10^{7}\,{\rm s^{-1}}$ & decay from $\ket{8}$ to $\ket{5}$ \\ 
2  & $\sig_{68}$   & $\gamma_{86}$ & $1.68\times10^{7}\,{\rm s^{-1}}$ & decay from $\ket{8}$ to $\ket{6}$ \\ 
3  & $\sig_{68}$   & $\gamma_{87}$ & $1.26\times10^{7}\,{\rm s^{-1}}$ & decay from $\ket{8}$ to $\ket{6}$ \\ 
4  & $\sig_{48}$   & $\gamma_{84}$ & $1.35\times10^{7}\,{\rm s^{-1}}$ & decay from $\ket{8}$ to $\ket{4}$ \\ 
5  & $\sig_{38}$   & $\gamma_{83}$ & $1.35\times10^{7}\,{\rm s^{-1}}$ & decay from $\ket{8}$ to $\ket{3}$ \\ 
6  & $\sig_{e8}$   & $\gamma_{8e}$ & $8.9\times10^{6}\,{\rm s^{-1}}$ & decay from $\ket{8}$ to $\ket{e}$ \\ 
7  & $\sig_{g4}$   & $\gamma_{4g}$ & $4.69\times10^{7}\,{\rm s^{-1}}$ & decay from $\ket{4}$ to $\ket{g}$ \\ 
8  & $\sig_{g3}$   & $\gamma_{3g}$ & $4.69\times10^{7}\,{\rm s^{-1}}$ & decay from $\ket{3}$ to $\ket{g}$ \\ 
9  & $\sig_{ge}$   & $\gamma_{eg}$ & $\gamma$  & decay from $\ket{e}$ to $\ket{g}$ \\ 
\hline
10  & $\sum_{k=3}^8\sig_{kk}$   & $\Gamma^L_{1}$ & about kHz & Fluctuations of laser acting on $\ket{g} \rightarrow \ket{3}$ transition \\ 
11  & $\sum_{k=4}^8\sig_{kk}$   & $\Gamma^L_{3}$ & about MHz & Fluctuations of laser acting on $\ket{3} \rightarrow \ket{8}$ transition \\ 
12  & $ \sig_{44}$   & $\Gamma^L_{4}$ & about MHz & Fluctuations of laser acting on $\ket{4} \rightarrow \ket{8}$ transition \\ 
13  & $ \sig_{55}$   & $\Gamma^L_{5}$ & about MHz & Fluctuations of laser acting on $\ket{5} \rightarrow \ket{8}$ transition \\ 
14  & $ \sig_{66}$   & $\Gamma^L_{6}$ & about MHz & Fluctuations of laser acting on $\ket{6} \rightarrow \ket{8}$ transition \\ 
15  & $ \sig_{77}$   & $\Gamma^L_{7}$ & about MHz & Fluctuations of laser acting on $\ket{7} \rightarrow \ket{8}$ transition \\ 
\hline\hline
\end{tabular}
\label{table:JumpRates8level}
\end{table*}

As a next step we, following the approach used in \cite{Hotter20V} and switch into a instantaneous rotating frame with the unitary transformation:
\begin{widetext}
\begin{equation}
\begin{split}
\hat{U}=\exp \Big[ -i \Big( &
    \sig_{gg}\, \omega_{g}t +\sig_{ee}\, \omega_{e}t + \sig_{33}\, [(\omega_g +\omega_{3g}^L)t+\phi_{1}] 
    + \sig_{88} [(\omega_g+\omega_{3g}^L+\omega_{83}^L)t+\phi_{13}+\phi_{3}] 
\\
+ &  \sum_{i=4}^7
    \sig_{ii} [(\omega_1+\omega_{31}^L+\omega_{83}^L-\omega_{84}^L)t+\phi_{1}+\phi_{3}-\phi_{i}] 
\Big) \Big].
\end{split}
\label{eq:C:U}
\end{equation}
\end{widetext}

The new Hamiltonian
\begin{equation}
    \HH=\hat{U}^\dagger \HH^0 \hat{U} - i \hbar \hat{U}^\dagger \frac{\partial \hat{U}}{\partial t} = \HH^D + \sum_j \HH^S_j \dot{\phi}_j
\label{eq:C:Htrans}
\end{equation}
can be represented as a sum of deterministic part $\HH^D$ and a series of stochastic parts $\HH^S_i \dot{\phi}_i$. The deterministic part can be written as
\begin{equation}
    \frac{\HH^D}{\hbar}=\sum_{j=3}^8 \Delta_j \sig_{jj} + \Omega_1(\sig_{g3}+\sig_{3g})+\sum_{j=3}^7 \Omega_j (\sig_{j8}+\sig_{8j}),
\label{eq:C:HD}
\end{equation}
where 
\begin{equation}
\begin{split}
\Delta_3&=\omega_3-\omega_g-\omega^L_{3g}, \\
\Delta_4&=\omega_4-\omega_g-\omega^L_{3g}-\omega^L_{83}+\omega^L_{84}, \\
\Delta_5&=\omega_5-\omega_g-\omega^L_{3g}-\omega^L_{83}+\omega^L_{85}, \\
\Delta_6&=\omega_5-\omega_g-\omega^L_{3g}-\omega^L_{83}+\omega^L_{86}, \\
\Delta_7&=\omega_5-\omega_g-\omega^L_{3g}-\omega^L_{83}+\omega^L_{87}, \\
\Delta_8&=\omega_8-\omega_g-\omega^L_{3g} -\omega^L_{83}. 
\end{split}
\label{eq:C:Detunings}
\end{equation}
In turn, stochastic parts has the form

\begin{equation}
\begin{split}
\frac{\HH^S_1}{\hbar}=-\sum_{j=3}^8 \sig_{jj}, &\quad
\frac{\HH^S_3}{\hbar}=-\sum_{j=4}^8 \sig_{jj}, \\
\frac{\HH^S_4}{\hbar}= \sig_{44}, &\quad
\frac{\HH^S_5}{\hbar}= \sig_{55}, \\
\frac{\HH^S_6}{\hbar}= \sig_{66}, &\quad
\frac{\HH^S_7}{\hbar}= \sig_{77}.
\end{split}
\label{eq:C:HS}
\end{equation}

Evolution of some system operator $\oo$ can be described by the Langevin-Heisenberg equation
\begin{equation}
({\rm S})\frac{d\oo}{dt}=\frac{i}{\hbar}\left[\HH^D,\oo\right]+\LL_{\rm dec}[\oo] + \sum_j \frac{i}{\hbar}\left[\HH^S_j,\oo\right]\dot{\phi}_j,
\label{eq:C:ME:Stra}
\end{equation}
which needs to be interpreted as a Stratonovic stochastic differential equation (indicated by (S)). Here $\LL_{\rm dec}[\oo]$ is a Liouvillian term describing spontaneous transitions between atomic levels 
\begin{equation}
\LL_{\rm dec}[\oo] = \sum_{k,l}\frac{\gamma_{kl}}{2}
\left(2\sig_{kl}\oo\sig_{lk}-\sig_{kk}\oo-\oo\sig_{kk}\right)
\label{eq:C:Dec}
\end{equation}
where $\gamma_{kl}$ is the spontaneous transition rate $\ket{k} \rightarrow \ket{l}$.

Equation (\ref{eq:C:ME:Stra}) can be transformed into the Ito form (indicated by (I)) as
\begin{equation}
\begin{split}
({\rm I})\frac{d\oo}{dt}=&\frac{i}{\hbar}\left[\HH^D,\oo\right]+\LL_{\rm dec}[\oo] + \sum_j \frac{i}{\hbar}\left[\HH^S_j,\oo\right]\dot{\phi}_j\\
&+ \sum_j \frac{\Gamma^L_j}{2}\left(2\HH^S_j \oo \HH^S_j
-\HH^S_j \oo - \oo \HH^S_j \right),
\end{split}
\label{eq:C:ME:Ito}
\end{equation}
this approach has also been used in Ref \cite{Hotter20V}. Here, we used ${\HH^{S\dagger}_j}=\HH^S_j=\HH^{S2}_j$. By averaging this equation the stochastic part vanishes, and we get
\begin{equation}
\frac{d}{dt}\langle\oo\rangle=\frac{i}{\hbar}\left\langle\left[\HH^D,\oo\right]\right\rangle+\langle\LL[\oo] \rangle.
\label{eq:C:ME}
\end{equation}
The dissipative processes are described by the Liouvillian part
\begin{equation}
\LL[\oo]=\LL_{\rm dec} [\oo] + \sum_j \frac{\Gamma^L_j}{2}\left(2\HH^S_j \oo \HH^S_j
-\HH^S_j \oo - \oo \HH^S_j \right)
\label{eq:C:Liouv1}
\end{equation}
which can be represented as
\begin{equation}
\LL[\oo]=\sum_j \frac{R_j}{2}\left( 2 \hat{J}_j^\dagger \oo \hat{J}_j-\hat{J}_j^\dagger \hat{J}_j\oo-\oo\hat{J}_j^\dagger \hat{J}_j\right),
\label{eq:C:Liouv2}
\end{equation}
where $\hat{J}_j$ are {\em jump operators} with corresponding rates $R_j$. The list of jump operators and rates for the full 8-level system is presented in Table~\ref{table:JumpRates8level}

In equivalent 2-level system, an averaged value of the operator $\oo$ can be expressed in the equation as the following form (\ref{eq:C:Liouv2}), where the (deterministic) Hamiltonian is equal to
\begin{equation}
\HH^D_{\rm 2-level}=\hbar (\delta_g \sig_{gg}+\delta_e \sig_{ee}),
\end{equation}
and the dissipative processes are listed in Table~\ref{table:JumpRates2level}.
\begin{table}
\caption{Dissipative processes in the equivalent 2-level scheme with incoherent pumping.}
\centering
\begin{tabular}{cccc}
\hline\hline
$j$& $\hat{J}_j$   & $R_j$         & description \\ \hline
1  & $\sig_{ge}$   & $\gamma$      & decay from $\ket{e}$ to $\ket{g}$ \\ 
2  & $\sig_{eg}$   & $w$           & incoherent pumping from $\ket{g}$ to $\ket{e}$ \\ 
3  & $\sig_{gg}$   & $\nu_{g}$     & dephasing on $\ket{g}$ \\ 
4  & $\sig_{ee}$   & $\nu_{e}$     & dephasing on $\ket{e}$ \\ 
\hline\hline
\end{tabular}
\label{table:JumpRates2level}
\end{table}

\newcommand{\STAB}[1]{\begin{tabular}{@{}c@{}}#1\end{tabular}}
\begin{table*}
\caption{The pumping parameters of the realistic 8-level scheme and equivalent 2-level scheme. Here we introduced $\Delta \omega_{kl}=\omega^{L}_{kl}-\omega_k+\omega_l$ and supposed that all the lasers except the one acting on $\ket{g} \rightarrow \ket{e}$ transition have the same linewidth. Here $\nu=\nu_g+\nu_e$, and $\delta_p=\delta_e-\delta_g$.}
\centering
\begin{tabular}{cccccc}
\hline\hline
   & \multicolumn{1}{c}{Parameters} & \multicolumn{1}{c}{Values: 1st Set} & \multicolumn{1}{c}{Values: 2nd Set} & \multicolumn{1}{c}{Values: 3rd Set}
   & \multicolumn{1}{c}{Values: 4th Set}\\ \hline
\multirow{8}{*}{\STAB{\rotatebox[origin=c]{90}{8-level scheme}}}  
& $\Delta \omega_{3g}/(2\pi)$  & $ 20$ Hz   & $-100$ Hz   & $ 50$ Hz   & $ 500$ Hz   \\ 
& $\Delta \omega_{83}/(2\pi)$  & $-1$ MHz   & $-2.0$ MHz  &  0         & 0 \\ 
& $\Delta \omega_{84}/(2\pi)$  & $1$ MHz    &  0          & $ 2$ MHz   & $2$ MHz \\ 
& $\Delta \omega_{85}/(2\pi)$  &$-20$ MHz   & $-19.9$ MHz &  $-20$ MHz & $-17$ MHz \\  
& $\Delta \omega_{86}/(2\pi)$  &  0         & $- 100$ kHz &  0         & $- 3$ MHz \\ 
& $\Delta \omega_{87}/(2\pi)$  & $20$ MHz   & $ 20.1$ MHz &  $20$ MHz  & $-23$ MHz 
\\ 
& $\Gamma^L_{1}/(2\pi)$        & $1$ kHz    & $1  $ kHz   &  $1$ kHz   & $2$ kHz \\  
& $\Gamma^L_{j}/(2\pi),~j=3~{\rm to}~8$ & $ 3$ MHz & $3$ MHz &  $1$ MHz & $ 1$ MHz \\  
\hline
\multirow{3}{*}{\STAB{\rotatebox[origin=c]{90}{2-level }}}  
& $w$                   & $272.6~{\rm s^{-1}}$   & $272.8~{\rm s^{-1}}$     & $248.0~{\rm s^{-1}}$  & $246~{\rm s^{-1}}$ \\ 
& $\nu=\nu_g+\nu_e$     & $401~{\rm s^{-1}} $    & $404~{\rm s^{-1}} $      & $346~{\rm s^{-1}} $   & $341~{\rm s^{-1}} $ \\ 
& $\delta_p/(2\pi)$ 
                  & $ 6.26~{\rm Hz}$ & $12.4~{\rm Hz}$ & $0.02~{\rm Hz}$ & $0.42~{\rm Hz}$ \\ 
\hline\hline
\end{tabular}
\label{table:EquivParameters}
\end{table*}

The procedure of mapping of multilevel pumping scheme to an equivalent 2-level scheme with incoherent pumping is described in detail in \cite{Hotter21repump}. In brief, as a first step one has to find the steady-state values of $\langle\sig_{ee}\rangle$ and $\langle\sig_{gg}\rangle$, solving the master equation (\ref{eq:C:ME}). Then the equivalent incoherent pumping rate is
\begin{equation}
w=\gamma_{eg} \frac{\langle\sig_{ee}\rangle}{\langle\sig_{gg}\rangle}.
\label{eq:C:w}
\end{equation}
Second, one has to diagonalise the effective non-hermitian Hamiltonian which is expressed here:
\begin{equation}
\HH_{\rm eff}^{\rm nh}=\HH^D-\frac{i\hbar}{2}\sum_jR_j \hat{J}^+_j\hat{J}_j
\label{eq:C:Heff}
\end{equation}
to get the complex eigenvalues $E_g$ and $E_e$, corresponding to the eigenstates with the highest overlap with the unperturbed ``clock'' states $\ket{g}$ and $\ket{e}$. Then one can extract the effective frequency shifts
\begin{equation}
\delta_{e,g}=\Re(E_{e,g})
\label{eq:C:Shifts}
\end{equation}
and dephasing rates
\begin{equation}
\nu_{g}=-2\Im(E_g)-\gamma_{eg}; \quad \nu_{e}=-2\Im(E_e)-w.
\label{eq:C:Dephasings}
\end{equation}

In Table~\ref{table:EquivParameters} we present four examples of mapping the realistic 8-level pumping scheme into the effective 2-level scheme using incoherent pumping. For all four sets of parameters Appendix~\ref{App:B}, $I=0.32\,I_{\rm sat}$ is the laser acting on $\ket{g}\rightarrow \ket{3}$ transition that gives $\Omega_1=9.38\times 10^3~{\rm s^{-1}}$), $I=0.5\,I_{\rm sat}$ for laser acting on $\ket{3}\rightarrow \ket{8}$ transition, what gives $\Omega_3=3.375\times 10^6\,{\rm s^{-1}}$ and $I=0.02\, I_{\rm sat}$ for all other lasers, that gives $\Omega_{3}=\Omega_{4}=6.75\times 10^5~{\rm s^{-1}}$, $\Omega_{5}=\Omega_{7}=6.3\times 10^5~{\rm s^{-1}}$, $\Omega_{6}=8.4\times 10^5~{\rm s^{-1}}$, according to expression (\ref{eq:C:OmegaVsIsat}).

Finally, let us evaluate the possible influence of the scattered photons during the pumping process on the clock transition inside the emission zone. We have estimated that each atom scatters, on average, about 12 photons in total, and about 2.76 of them are 689~nm photons. The 689~nm photons are of most interesting because they can affect the $^1S_0$ state and so disturb the clock transition. Taking $\Phi=10^7~{\rm 1/s}$, we can estimate the intensity $I_{\rm scat}$ of scattered photons is as about $I_{\rm scat}\approx 2\times 10^{-5}I_{\rm sat}$ at a distance of 1~mm from the pumping zone. The resulting light shift $\Delta_{\rm LS,scat}$ for such a small intensity can be estimated as
\begin{equation}
|\Delta_{\rm LS,scat}|=\frac{I_{\rm scat}\gamma_{3g}^2|\Delta_3|}{2I_{\rm sat}(\gamma_{3g}^2+4\Delta_3^2)} < \frac{I_{\rm scat}}{8I_{\rm sat}}\gamma_{3g},
\label{eq:appC:scat}
\end{equation}
which results in light shift of $|\Delta_{\rm LS,scat}|<2\pi \times 20~{\rm mHz}$. 

Similarly, using expression (\ref{app:sr}), we can estimate the ``rescattering rate'' $\Gamma_{\rm rescat}$ of the scattered photons on $\ket{g}\rightarrow\ket{3}$ transition as $\Gamma_{\rm scat,2}<\gamma_{3g}I_{\rm scat}/(8I_{\rm sat})\approx0.13~{\rm s^{-1}}$. These effects can be neglected because it even smaller than the previous value. It need to be noted that the photons scattered from the molasses beams produces a similar light shift but the ``rescattering rate'' is two orders of magnitude smaller due to differential light shift between reservoir and conveyor lattice. On average every atoms scatters around 300 photons during the cooling process in the reservoir but these photons are detuned by about $2\pi \times 1~{\rm MHz}$ from the $\ket{g}\rightarrow\ket{3}$ transition of the atom in emission zone, see Table~\ref{table:PMB}. Therefore, both light shifts caused by photons scattered during the cooling and pumping processes, as well as dephasing due to rescattering, are very small in comparison to the other considered effects and can be neglected.

\section{Details of simulation of the superradiant laser}
\label{App:D}

Here, we present details of the atom number density calculation in the conveyor that we used for further calculations of collision-induced losses and shifts in (\ref{eq:conv:8}) -- (\ref{eq:conv:12}) and (\ref{eq:conv:23}). Also, it is used to determine the coupling strength $g$ between the atom and the field. Next, the periods of the atom radial and axial motion in a single lattice site (of order of $0.01~{\rm s}$ and $10^{-5}~{\rm s}$, respectively) are much shorter than the interaction time between the atom and the field (around 1~s). Therefore, we can average position-dependent terms and spatial distribution of the atoms in the lattice site. To perform this averaging, we used a harmonic oscillator approximation for the dipole potential of the lattice site and a Maxwell-Boltzmann spatial distribution of the atoms:
\begin{equation}
p(x',y',z')=\frac{2^{3/2}}{\pi^{3/2}W_r^2 W_y} \,
\exp\left(-2\frac{x'^2+z'^2}{W_r^2}-2\frac{y'^2}{W_y^2} \right),
\label{eq:D:1}
\end{equation}
where $x',y'$ and $z'$ are distances from the center of the lattice site. The $1/e^2$ radii $W_r$ and $W_y$ of the atomic cloud in the radial and axial directions are calculated as
\begin{align}
W_r&=W_{\rm conv} \sqrt{T/\Uconv}
\label{eq:D:2} \\
W_y&=\frac{1}{k}\sqrt{2\frac{(T^\theta+(\Uconv \Erec)^{\theta/2})^{1/\theta}}{\Uconv}},
\label{eq:D:3}
\end{align}
where $T$ is the temperature of the atomic ensemble set to  $T=10~{\rm \mu K}$. Next, $\Uconv =30~{\rm \mu K}$ is the depth, and $\Erec = k^2 \hbar^2/(2 \mSr k_B) \approx 0.165~{\rm \mu K}$ is the recoil energy of the moving optical lattice potential in units of temperature.
The phenomenological parameter $\theta=2.5$ is chosen such that Eq.~(\ref{eq:conv:22}) well reproduces the probability density of the atom in the harmonic potential in the cross-over between the ``classical thermal'' limit ($k_B T \gg \hbar \omega_{\rm y}=2\sqrt{\Erec \Uconv}$) and the ``frozen quantum'' limit ($k_B T \ll \hbar \, \omega_y$). Here we consider the possibility that higher vibrational states along the $y$-axis can be occupied.
The number density $n_j$ averaged over the atomic motion can be represented as
\begin{align}
n_j=N_{\rm } \int p^2(x',y',z')dx'dy'dz'=\frac{N_{\rm lc}}{\Veff},
\label{eq:D:4}
\end{align}
where $N_{\rm lc}=N^j\lambda_{\rm conv}/(2\lc)$ is the number of atoms in a single lattice site, and

\begin{equation}
\Veff=W_r^2 W_y \pi^{3/2}
\label{eq:D:5}
\end{equation}
is the effective volume of a single lattice site. 

The atom - cavity coupling coefficient is then 
\begin{equation}
g=\frac{\exp\left( \frac{-k^2 W_y^2}{8}\right)}{1+\frac{W_r^2}{2W_0^2}} \sqrt{\frac{3 c^3 \gamma}{\lcav \omega^2 W_0^2}} ,
\label{eq:D:6}
\end{equation}
where the prefactor before the square root describes the averaging over the spatial distribution of the atoms in a single lattice site, and $k=2\pi/\lambda_{\rm conv}$ is the wave number of the conveyor lattice. 

Finally, a short description for the shift coefficients $\mu$ and $\epsilon$. In \cite{Lisdat09}, the collision shift coefficient was measured as $(7.2 \pm 2.0)\times 10^{-17}~{\rm Hz \times m^3}$ for $35\%$ of excited atom in the end of the Rabi pulse, that averages to about $31.75\%$ of excited atoms per pulse. To calculate $\mu$ and  $\epsilon$, we converted percentage of excited atoms after the pulse into percentage of excited atoms averaged over the pulse.

\bibliography{main}
\bibliographystyle{unsrt}
\end{document}